\documentclass[12pt,preprint]{aastex}
\usepackage{mathrsfs}
\begin{document}
\title{Early phase of massive star formation: A case study of Infrared dark cloud G084.81$-$01.09}
\author{S. B. Zhang\altaffilmark{1}, J. Yang\altaffilmark{1}, Y. Xu\altaffilmark{1},  J. D. Pandian\altaffilmark{2}, K. M. Menten\altaffilmark{3}, C. Henkel\altaffilmark{3}}

\altaffiltext{1}{Purple Mountain Observatory, Chinese Academy of Sciences, Nanjing 210008, China; shbzhang@pmo.ac.cn}
\altaffiltext{2}{Institute for Astronomy, 2680 Woodlawn Drive, Honolulu, HI 96822, USA} \altaffiltext{3}{Max-Planck-Institut f$\ddot{u}$r Radioastronomie, Auf dem H$\ddot{u}$gel 69, 53121 Bonn,
Germany}
\begin{abstract}

We mapped the MSX dark cloud G084.81$-$01.09 in the NH$_3$~(1,1) - (4,4) lines and in the $J$ = 1-0 transitions of $^{12}$CO, $^{13}$CO, C$^{18}$O and HCO$^+$ in order to study the physical properties of infrared dark clouds, and to better understand the initial conditions for massive star formation. Six ammonia cores are identified with masses ranging from 60 to 250~$M_\sun$, a kinetic temperature of 12~K, and a molecular hydrogen number density $n({\rm H_2})\sim10^5$~cm$^{-3}$. In our high mass cores, the ammonia line width of 1~km~s$^{-1}$ is larger than those found in lower mass cores but narrower than the more evolved massive ones. We detected self-reversed profiles in HCO$^+$ across the northern part of our cloud and velocity gradients in different molecules. These indicate an expanding motion in the outer layer and more complex motions of the clumps more inside our cloud. We also discuss the millimeter wave continuum from the dust. These properties indicate that our cloud is a potential site of massive star formation but is still in a very early evolutionary stage.

\end{abstract}

\keywords{stars: formation -- ISM: molecules -- ISM: kinematics and dynamics -- ISM: individual (G084.81$-$01.09)}

\section{INTRODUCTION}

Infrared dark clouds (IRDCs) are identified as small areas with high extinction against the bright diffuse mid-infrared background along the Galactic Plane. These ``flecks'' in the sky were first discovered by the 15~$\mu$m ISOGAL survey (the inner Galactic disk survey of the Infrared Space Observatory, \citealt{per96}) and later cataloged and studied using the 8.3~$\mu$m survey data of the Mid-course Space Experiment (MSX, \citealt{ega98}). \citeauthor{ega98} found about 2000 IRDCs and note that these objects are cold and dense molecular cores. These results were refined by later molecular line observations \citep{car98,tey02} and radio continuum observations \citep{car00}. These authors concluded that these clouds have low kinetic temperatures of 10-20~K and a gas density over $10^5$~cm$^{-3}$. Their small line widths of 1 - 3~km~s$^{-1}$ can be related to a very early phase of star formation. Therefore, the IRDCs are good candidates for pre-protostellar cores and may be the primary sites for future massive star formation, which plays an important role in the evolution of our galaxy.

However, IRDCs remain mysterious in at least some aspects. A controversial issue being discussed is whether the cloud evolution is quasi-static or whether it is undergoing a dynamical process caused by lack of equilibrium \citep{ber07}. Understanding the role of supersonic turbulence in dark clouds is important in settling this argument. Evidence from current observations of IRDCs seems to support both views. Most IRDC cores are supported by turbulent pressure but appear to be virialised \citep{pil06}. Furthermore, the origin of the turbulence in the molecular clouds and the transition from the turbulent cloud to the quiescent core regime is still unclear. Also, more systematic studies need to be done to compare them with other relatively nearby massive star-forming clouds. The answers will provide a critical element in our general understanding of star formation.

In this study, we have obtained spectroscopic data in several molecules towards the dark cloud G048.81--01.09, a source in the catalog of \citet{sim06}. The cloud has an elongated morphology (N-S) with an extent of $\sim 15$ arcmin, and is situated in a region of high extinction ($A_v > 20$, \citealt{cam02}) of the dark cloud LDN935 \citep{lyn62} which separates the W80 HII region into the North America (NGC 7000) and Pelican (IC 5070) nebulae \citep{com05}. \citet{com05} proposed that the ionizing source of the complex is 2MASS~J205551.25+435224.6, an O5V star located behind LDN935. The IRDC is almost starless with no associated IRAS or MSX point sources. Previous molecular line observations indicated the presence of large amounts of gas with different velocity components and signs of ongoing star formation in LDN935 around our dark cloud \citep{fel93}. A single pointed CO observation towards our cloud conducted by \citet{mil75} suggests however that the gas is not appreciably heated.

There is no direct distance measurement for G084.81$-$01.09, but it is associated with dark cloud LDN935 for which there are a number of distance estimates. A radio continuum study carried out by \citet{wen83} gave a distance estimate of 500~pc. \citet{str93} obtained a consistent result of 580~pc on the basis of photometric results of 564 stars toward three areas near the North America and Pelican Nebulae complexes. New measurements done by \citet{cer07} derived a distance of about 0.7~kpc, which was obtained from radio recombination line observations at a frequency around 1.4~GHz and by applying a flat rotation-curve model. \citet{com05} obtained a distance of 610~pc to the ionizing star, 2MASS~J205551.25+435224.6. Here we adopt a distance of 580~pc in our calculation of physical parameters.

\section{OBSERVATIONS AND DATA REDUCTION}\label{sec:obs}


\subsection{Effelsberg 100 m Observations}

We mapped the infrared dark cloud G084.81$-$01.09 in the NH$_3$~(1,1), (2,2), (3,3) and (4,4) transitions using the Effelsberg 100~m telescope of the Max-Planck-Institut f\"{u}r Radioastronomie (MPIfR) during December 12 - 14, 2007 (see Table~\ref{tbl:observation} for adopted line frequencies). The spectrometer backend employed was the 8192 channel AK90 auto-correlator which consists of 8 individual 1024 channel correlators of 20~MHz bandwidth each. Additional data of NH$_3$~(1,1), (2,2) and (3,3) were obtained on February 21 - 24, 2008 using the Fast Fourier Transform Spectrometer (FFTS), which has 16382 channels covering a bandwidth of 500~MHz. The data were taken in using frequency switching with a frequency throw of 7.5~MHz. The map covers the area with notable extinction in the MSX 8.3~$\mu$m image with a grid spacing of 30\arcsec. The pointing was checked roughly every hour by observations of nearby continuum sources and was found to be accurate to better than 10\arcsec.

Continuum scans were used for calibrating the absolute flux density. The raw data were scaled from arbitrary units to main beam brightness temperature by applying gain-elevation corrections and the flux scale was set using NGC~7027 by assuming it to have a flux density of 5.4~Jy at our observing frequency \citet{ott94}. Our absolute calibration uncertainty is $\pm$10\%.

\subsection{PMODLH 13.7 m Observations}

The $^{12}$CO~(1$-$0), $^{13}$CO~(1$-$0), C$^{18}$O~(1$-$0) and HCO$^+$~(1$-$0) observations of the MSXDC G084.81$-$01.09 (line frequencies shown in Table~\ref{tbl:observation}) were carried out at the Purple Mountain Observatory Delingha (PMODLH) 13.7~m telescope from March to June in 2008. The 3~mm SIS receiver was used in double sideband mode. Using three 1024 channel AOS spectrometers with bandwidths of 145.43~MHz, 42.87~MHz and 43.3~MHz, the three CO lines were observed simultaneously. The source was mapped using position switched observations, with the standard chopper wheel method for calibration \citep{pen73}. Standard sources were checked roughly every 2 hours.

To compare with the NH$_3$ data, our CO map covered the area mapped with NH$_3$ and extended a few arcmins to the northeast. The grid spacing was 30\arcsec\ in the center and 60\arcsec\ in other regions. The HCO$^+$ map covered a region similar to that of the NH$_3$ map with a typical grid spacing of 30\arcsec.

A summary of all observation parameters is provided in Table~\ref{tbl:observation}.

\subsection{Data Reduction}

We used the CLASS software package\footnote{CLASS is part of the GILDAS software suite available from \url{http://www.iram.fr/IRAMFR/GILDAS}} for the spectral data reduction. After discarding the bad scans, the spectra of each position were averaged and a polynomial baseline of orders 1-3 was subtracted. Bad channels were excluded from the baseline fitting.

The method ``NH3(1,1)'' in CLASS \citep{bui05} was used to fit the hyperfine structure of the NH$_3$~(1,1) spectra and to derive optical depths and line widths. The (2,2) line is so weak that only the main line could be dectected towards a few positions. Therefore, we simply fitted the (2,2) line with a single Gaussian to derive its main beam brightness temperature and an optical depth was not calculated in this case. Table~\ref{tbl:nh3 result} summarizes the fit parameters toward the cores detected in the data. No NH$_3$~(3,3) or (4,4) lines were detected in our observations.

The excitation temperature, rotational temperature, kinetic temperature and ammonia column density toward the cores are listed in Table~\ref{tbl:nh3 property}. These physical parameters were derived using the standard formulae for NH$_3$ spectra \citep{ho83}.

The CO and HCO$^+$ main beam brightness temperatures were derived from the antenna temperature, $T_A^*$, using a main beam efficiency, $\eta_{\rm mb}$, of 61\%. A first order baseline was applied for the CO spectra and 1st-3rd order baselines were used for the HCO$^+$ spectra. A single Gaussian fit to the spectra provided the line velocity with respect to the local standard of rest (LSR) and the line width for later discussion. The fit results are given in Table~\ref{tbl:co result}.

\section{RESULTS}\label{sec:res}

\subsection{Identification and Morphology of the Cloud}\label{sec:ide}

Fig.~\ref{fig:NH3+MIPS1} shows the NH$_3$ integrated intensity maps created from the velocity range -2.5 to 4.5 km~s$^{-1}$ (which excludes the satellite lines). Our images of ammonia emission toward G084.81$-$01.09 reveal extended, filamentary molecular emission that closely matches the morphology of the Spitzer mid-infrared extinction. Also, the ammonia maps (both NH$_3$~(1,1) and (2,2)) allow us to identify two main molecular condensations: one in the northeast, containing ammonia cores 1, 2 and 6, and one in the south, consisting of cores 3, 4 and 5. All cores defined in the NH$_3$~(1,1) map follow the standards given by \citet{tac00} and the name of the cores were designated in descending order of their integrated intensities, as tabulated in Table~\ref{tbl:nh3 result}. The NH$_3$~(1,1) core sizes range from 0.25 to 0.42~pc.

The NH$_3$ molecular emission also matches the 1.1~mm dust emission of Bolocam Galactic Plane Survey (BGPS, \citealt{agu10}) remarkably well as shown in Fig.~\ref{fig:MOL+CON}a. Furthermore, with the blue circles marking the clump peaks of continuum emission identified by \citet{ros09}, we note that the ammonia cores are coincident with the peaks in 1.1~mm emission.

The integrated intensity maps of $^{13}$CO~(1$-$0) and C$^{18}$O~(1$-$0) are shown in Fig.~\ref{fig:MOL+CON}b and \ref{fig:MOL+CON}c. The $^{13}$CO and C$^{18}$O emission lines show an extended feature along north-south direction, with a sharp cutoff toward the northwest. The extent of C$^{18}$O emission is larger than that of both the ammonia and continuum emission. The $^{13}$CO line, which tends to trace a lower gas density than C$^{18}$O, appears to be more extended. The strongest part of $^{13}$CO emission coincides with the northern condensation (peak at the position of ammonia core 1) and shows a tail to the east, while the $^{13}$CO emission in the southern part is flat with no peak associated with that in the C$^{18}$O map.

Fig.~\ref{fig:MOL+CON}d shows the integrated intensity of HCO$^+$~(1$-$0) overlaid on dust emission. The HCO$^+$ emission shows a different distribution compared to CO. Since the line is optically thick (see \S\ref{sec:lin}), it is likely to trace the envelope surrounding a dense core, while the C$^{18}$O line is expected to trace the dense core itself.

A candidate massive young stellar object (MYSO), G084.7847$-$01.1709, identified by \citet{urq09} in their 6~cm VLA survey is marked with cross symbol on Fig.~\ref{fig:NH3+MIPS1}. It is spatially coincident with the north condensation of our cloud. However, no compact source can be found near this object in the Spitzer 24 or 70~$\mu$m data. The lack of infrared emission could be due to the source being an extragalactic object rather than a Galactic MYSO. Further observations at centimeter wavelengths are needed to determine the spectral index of emission and the nature of this object.

\subsection{Line intensities and kinetic temperature}\label{sec:int}

The excitation temperature, $T_{\rm ex}$, of the NH$_3$~(1,1) transition (Column $T_{\rm ex}$ in Table~\ref{tbl:nh3 property}) was obtained from the optical depth via the relation
\begin{displaymath}
T_{\rm mb}={h\nu \over k}[J(T_{\rm ex})-J(T_{\rm bg})](1-e^{-\tau}), J(T)=(e^{h\nu \over kT}-1)^{-1},
\end{displaymath}
where $T_{\rm mb}$ and $\tau$ represent the temperature and the optical depth from CLASS fitting procedures and $T_{\rm bg}$ equals 2.7~K. The rotational temperature between the NH$_3$~(1,1) and (2,2) inversion doublets were then derived from the excitation temperature using the method given by \citet{ho83}. We then estimate the kinetic temperature using the expression of \citet{taf04}.

The typical kinetic temperature in the cores is about 12~K as seen in Table~\ref{tbl:nh3 property}. No obvious difference in temperature appears among the six cores. The kinetic temperature in the envelope is about 2~K higher than in the cores. \citet{des08} carried out a low resolution ($\sim$12 arcmin) survey at four (sub)millimeter wavelengths and derived a dust temperature of 8.5~K with a fixed emissivity law exponent of 2 toward our cloud, which is slightly lower than our gas temperature, probably because of the different beam size.

The physical parameters of CO are calculated under the assumption of Local Thermodynamic Equilibrium (LTE), in which the $^{13}$CO and C$^{18}$O lines are assumed to reach the same excitation temperature as $^{12}$CO line. The $^{12}$CO~(1-0) line is optically thick, so we can derive the kinetic temperature directly from its brightness temperature. The kinetic temperature is found to be 12 -- 14~K as listed in Table~\ref{tbl:co property} and agrees with that derived from ammonia.

The $^{13}$CO and C$^{18}$O line intensity ratio indicates the different abundances and optical depths in the cloud. As shown in Fig.~\ref{fig:TrCO}, the $^{13}$CO to C$^{18}$O ratios in regions with strong emission are low (about 1.4-2.5), while the ratios of faint emission areas are high and close to the local interstellar [$^{13}$CO]/[C$^{18}$O] ratio of about 7.3 \citep{wil94,tey02}. The observation of variable line ratios as a function of intensities is consistent with that of \citet{tey02}.

\subsection{Line widths and profiles}\label{sec:lin}

The NH$_3$~(1,1) line widths in the cloud vary from 0.4 to 2.8~km~s$^{-1}$, which is consistent with the value reported from massive dense cores \citep{ben89,pil06}. The distribution of ammonia line widths is shown in Fig.~\ref{fig:NH3 velocity}. The line widths in the peripheral regions ($>$1.3~km~s$^{-1}$) are generally larger than those in the clump centers (0.7 - 1.4~km~s$^{-1}$). The line width of Core 6 is influenced and broadened by the extended part of core 1 and double peaked spectra appear in the central part of core 6 (shown in Fig.~\ref{fig:spectra}, P6). For cores 2, 4, 5 and 6, the (2,2) line widths are larger than the (1,1) line widths, which suggests that the (2,2) line may not trace the same gas as the (1,1) line toward these cores \citep{pil06}.

As seen in Fig.~\ref{fig:spectra}, the CO lines show multiple velocity components. Most of the $^{13}$CO spectra in the cores are blended with a component around 5.5~km~s$^{-1}$, which also appears in some C$^{18}$O spectra. This component is consistent with the extended part of another cloud to the west of core 1, reported by \citet{fel93}. A few of the $^{13}$CO spectra are flat topped, indicating that the line is optically thick at these locations. As with the $^{13}$CO line profiles, the C$^{18}$O spectra are also slightly asymmetric, especially in core 2. The line widths over the core regions range between 2.3 and 4.6~km~s$^{-1}$ for $^{13}$CO and between 1.5 and 2.7~km~s$^{-1}$ for C$^{18}$O.

The thermal line width, $[(8 \ln 2) k T_{kin} / m]^{1/2}$ ($k$ is the Boltzmann constant and $m$ is the mean molecular mass), at a kinetic temperature, $T_{kin}$, of 12~K is 0.18~km~s$^{-1}$ for NH$_3$ and 0.14~km~s$^{-1}$ for $^{13}$CO and C$^{18}$O. Therefore, the non-thermal line widths are significantly larger than the thermal line widths and near the observed line widths. This suggests that non-thermal broadening mechanisms (rotation, turbulence, etc) play a dominant role in producing the observed line profiles.

The HCO$^+$~(1-0) spectra as shown in Fig.~\ref{fig:HCOP+C18O} display three types of profiles: an asymmetric, double-peaked shape in the northern C$^{18}$O condensation (e.g., B, C, P1, P2, etc.), spectra with a blue shoulder in the far north (e.g., A, P6, etc.) and a single blue-shifted peak in the south (e.g., D, etc.). The single peaked C$^{18}$O line appears to bisect the HCO$^+$ profiles in velocity, indicating that the HCO$^+$ lines are likely optically thick and self-absorbed.

The blue-shifted dip due to self-absorption may indicates that this occurs in an expanding envelope. By fitting Gaussian line profiles to these spectra and comparing the peak velocity of HCO$^+$ and C$^{18}$O spectra, one can roughly estimate an expansion velocity \citep{agu07}. We measure a velocity of $\delta V = |v_{\rm peak}({\rm C^{18}O})-v_{\rm peak}({\rm HCO^+})| \approx 1.05$~km~s$^{-1}$. This value is less than the C$^{18}$O line width at corresponding positions but is supersonic. On the contrary, the blue peaked profile in the south can be associated with infall asymmetry and may suggest infall motions of the outer cloud material. However, due to the low signal to noise ratio of the HCO$^+$ spectra nearby and the relatively large beam sizes involved, this motion can not be confirmed at all positions with blue-shifted peak. In order to better understand the spatial distribution of the HCO$^+$ line profile, we have compared the velocity of peak emission in the HCO$^+$ and C$^{18}$O lines for each location in the northern condensation, and mapped the velocity difference, $\delta V = v_{\rm peak}({\rm C^{18}O})-v_{\rm peak}({\rm HCO^+})$ (Fig.~\ref{fig:HCOP+C18O}). It can be seen that the outflow asymmetry (coded blue) dominates the central and northern regions of the cloud, while the infall asymmetry (coded red) appears on the southern edge.

\subsection{Velocity structure}\label{sec:vel}

From Table~\ref{tbl:nh3 property}, we note that the line width of core 2 is rather small (0.95~km~s$^{-1}$) but remarkably increases to 1.2~km~s$^{-1}$ when the spectra are averaged over the core area, which reflects the existence of velocity gradients in the core. This is also seen in the NH$_3$~(1,1) channel maps shown in Fig.~\ref{fig:NH3 channel} for velocities from $-$1.0 to 3.0~km s$^{-1}$ where different clumps appear at separate velocities. The clump structure is complex and covers a wide velocity range. In the lower velocity channels ($-$1.0 to 0.0~km~s$^{-1}$), the gas is concentrated in a clump in the east of northern condensation, while the higher velocity channels (0.5 to 2.0~km~s$^{-1}$) indicate the clump in the two peak position and a southern clump.

The CO presents multiple peaks along the line of sight (cf. CO spectra in Fig.~\ref{fig:spectra}). The channel maps of $^{13}$CO~(1-0) and C$^{18}$O~(1-0) (Fig.~\ref{fig:CO channel}) show that the different velocity components present various shapes and orientations. C$^{18}$O shows a velocity structure that is similar to that of ammonia, shown in Fig.~\ref{fig:NH3 channel}. The velocity increases from the position of core 1 at about 1~km~s$^{-1}$ to 2~km~s$^{-1}$ towards the north, which can be seen in position-velocity maps (Fig.~\ref{fig:pv map}) and the $^{13}$CO, C$^{18}$O and NH$_3$ channel maps.

To estimate the magnitude and direction of the velocity gradient in each core, we adopt the method of \citet{goo93} by fitting a linear velocity gradient to the projected velocity field along the line of sight. Thus the observed velocity $v_{\rm LSR}$ can be expressed as
\begin{displaymath}
v_{\rm LSR} = v_0 + \mathscr{G} \Delta\alpha \sin\Theta_\mathscr{G} + \mathscr{G} \Delta\delta \cos\Theta_\mathscr{G},
\end{displaymath}
where $\mathscr{G}$ is the magnitude of the velocity gradient, $\Delta\alpha$ and $\Delta\delta$ represent offsets in right ascension and declination in arcseconds, $v_0$ is the systemic velocity of the cloud, and $\Theta_\mathscr{G}$ is the direction of increasing velocity, measured east of north. We carry out a least-squares fit to the two dimensional velocity field of $^{13}$CO~(1-0), C$^{18}$O~(1-0), NH$_3$~(1,1) and NH$_3$~(2,2) weighted by $1/\sigma_{\rm LSR}^2$, where $\sigma_{\rm LSR}$ is the uncertainty in fitting $v_{\rm LSR}$. In Table~\ref{tbl:vgradient}, we list our fitting results ($\mathscr{G}$ and $\Theta_\mathscr{G}$) with their formal errors ($\sigma_\mathscr{G}$ and $\sigma_{\Theta_\mathscr{G}}$). We exclude fits that fail the $3\sigma$ criterion ($\mathscr{G}\geq3\sigma_\mathscr{G}$ mentioned by \citep{goo93}) or those with a random velocity field as seen from the velocity map.

Both cores 1 and 6 present significant velocity gradients in CO and NH$_3$ lines. The magnitude of the velocity gradient is greater in NH$_3$, and the direction of the gradient is somewhat different from that seen in CO. In cores 3, 4 and 5, the velocity gradient is detected only in NH$_3$~(1,1). The NH$_3$ gas associated with core 2 also exhibits a clear velocity gradient in Fig.~\ref{fig:NH3 velocity}. The peak lies on a velocity ridge with a gradient of 3.15~km~s$^{-1}$~pc$^{-1}$ to the east and 1.93~km~s$^{-1}$~pc$^{-1}$ to the west.

The distinct velocity gradients given by different molecules are probably due to the different densities they trace \citep{goo93}. Velocity gradients seen in the lower density tracers, $^{13}$CO or C$^{18}$O, probably arise in the envelope of the northern condensations. This gradient is dominated by unresolved clump-clump motions within the envelope, and may also be contaminated by the higher velocity component. The higher density tracers (e.g., NH$_3$), on the other hand, reveal the gradients within the small clumps themselves.

\subsection{Density and abundance}\label{sec:den}

The column densities of CO are estimated using the equation given by \citet{sco86}:
\begin{displaymath}
N={3k^2 \over 4h \pi^3 \mu^2 \nu^2} \exp\left({h \nu J \over k T_{\rm ex}}\right) \times {T_{\rm ex}+h \nu /6 k (J+1) \over {\rm exp}(-h \nu / k T_{\rm ex})} {\tau \over 1-{\rm exp}(-\tau)} \int{T_R^* d V},
\end{displaymath}
where $J$ is the rotational quantum number of the lower state in the observed transition, $\mu$ is the permanent dipole moment and $\tau$ is the optical depth derived from the observed ratios of $^{12}$CO and $^{13}$CO (or C$^{18}$O) emission. The derived physical properties are listed in Table~\ref{tbl:co property}. We also derive the column densities of NH$_3$ for the cores, which range from $15 - 34 \times10^{14}$~cm$^{-2}$ as shown in Table~\ref{tbl:nh3 property}. Since $^{13}$CO is not optically thin in some cores, we estimate the H$_2$ column densities based on the C$^{18}$O column densities where a factor $N({\rm H_2})/N({\rm C^{18}O})=7 \times 10^6$ was adopted \citep{fre82,kra99}. This gives H$_2$ column densities around $\sim10^{22}$ cm$^{-2}$ as listed in column (2) of Table~\ref{tbl:abundance}.

With the H$_2$ column densities derived from CO emission, we determine the ammonia abundances and list them in Table~\ref{tbl:abundance}. The ammonia abundance in the six peak positions is $3 - 5.5\times 10^{-8}$, which agrees with the parameters reported by \citet{pil06}, though their abundances are derived from the dust emission of SCUBA. When averaged over the cores, the abundances are reduced to $2.3 \times 10^{-8}$. The ammonia abundance and the correlation of its spatial distribution with the morphology of the dust emission is consistent with a chemical model given by \citet{ber97}, in which NH$_3$ is not depleted. We then compare its abundance with that of $^{13}$CO toward cores in Fig.~\ref{fig:relative abundance}. They differ slightly from each other with core 1 showing the highest abundance among the cores. This reveals the complex chemical environment in the cloud.

We obtain the gas mass of cores by integrating the total H$_2$ column density over the cores and show the results in Table~\ref{tbl:core}. The mass range from 60~$M_\sun$ to 250~$M_\sun$. The gas density can be derived by assuming the cores to be spherical and dividing the mass by the volume of the cores. An average density of $\sim1.1\times10^5$~cm$^{-3}$ was derived as shown in Table~\ref{tbl:core}.

The total H$_2$ column density can also be related to the 1.1~mm flux by assuming the continuum emission to be optically thin. After convolving the data to the resolution of the 100~m Effelsberg (40\arcsec) and 13.7~m Delingha (60\arcsec) beams, the dust based column density is given by
\begin{displaymath}
N({\rm H_2})_{\rm 1.1~mm}={S_{\rm 1.1~mm} \over \Omega_{\rm beam}B_\nu(T_d)\kappa_{\rm 1.1~mm}m},
\end{displaymath}
where $\Omega_{\rm beam}$ is the telescope beam solid angle, $B_\nu(T_d)$ is the Planck function at the dust temperature $T_d$ (assumed to be equal to the gas temperature deduced from NH$_3$), $\kappa_{\rm 1.1~mm}$ is the dust opacity at 1.1~mm, and $m$ is the mean molecular mass. Here we adopt $\kappa_{\rm 1.1~mm}$ to be $0.0114$~cm$^2$~g$^{-1}$ \citep{oss94,eno08} and assume a gas to dust mass ratio of 100. The results have the same magnitude as those derived from C$^{18}$O and are listed in Table~\ref{tbl:abundance}. We also compare our molecular emission with the BGPS survey, as shown in Fig.~\ref{fig:molevsdust}. The integrated intensity of NH$_3$~(1,1) shows a better correlation with the 1.1~mm dust emission compared to C$^{18}$O as mentioned previously.

The diversity in column densities derived from molecular gas and dust may be due to different structures and components they trace. The CO column density, because of the low density it traces, is highly affected by the structure of the cloud along the line of sight, especially at the positions where clumps with similar velocities blend together. Also the molecular observations are affected by high opacity or chemical depletion at high densities. On the other hand, calculations related to the dust emission are based on the value of dust opacity and assumptions about the dust to gas ratio. The dust opacity changes between different dust models and in environments of different density \citep{oss94}.

\section{DISCUSSION}\label{sec:ana}

\subsection{Evidence of massive star formation}

In this section, we will compare our results with those from other low-mass star forming regions, and discuss the evidence for massive star formation and the evolutionary state of the cloud. The mean intrinsic line width for NH$_3$~(1,1) in the samples given by \citet{jij99} (most of the objects are low-mass cores) is 0.5~km~s$^{-1}$. A recent work by \citet{fos09} also reported a small observed line width of 0.3~km~s$^{-1}$ for low mass NH$_3$ cores in the Perseus Molecular cloud. \citet{cra05} undertook a survey toward low-mass starless cores and also reported a narrow line width of 0.5~km~s$^{-1}$ for C$^{18}$O~(1-0). A broader line width of  1.2~km~s$^{-1}$ for C$^{18}$O~(1-0) is derived from a JCMT observation by \citet{jor02}, which is still less than those given in our observed region. Meanwhile, the dominant non-thermal line widths in our cloud indicate more non-thermal support than those in low-mass star forming regions. Although line widths can be broadened by active outflows in low-mass protostellar regions to 2~km~s$^{-1}$ for C$^{18}$O~(1-0) \citep{swi08} and 1.2~km~s$^{-1}$ for NH$_3$~(1,1 ) \citep{rud01}, at this stage, an active outflow has not developed yet. The NH$_3$ line widths here are comparable to those IRDCs of \citet{pil06}. Moreover, our derived column densities of NH$_3$, a few times 10$^{15}$~cm$^{-2}$, are higher than those of the low-mass cores ($\sim10^{13}-10^{15}$~cm$^{-2}$) given by \citet{suz92}, and are also comparable to those of the massive IRDCs given by \citet{pil06}. In addition, the supersonic motions of the envelope and the large velocity gradients in our cloud also indicate a more dynamic motion compared to low-mass cores. Therefore, we suggest that MSXDC G084.81$-$01.09 is potentially at an early evolutionary state of massive star formation.

\subsection{Gravitational stability of the cores}\label{sec:sta}

The expanding envelope detected from the line profiles described in \S\ref{sec:lin} lead us to consider the gravitational stability of MSXDC G084.81$-$01.09. By using the molecular mass, $M_{Molecular}$, derived from C$^{18}$O~(1-0) and the core radius $R$ determined from the NH$_3$~(1,1) maps, we obtain an escape velocity ($\sqrt{2GM/R}$) which ranges from $\sim 1.9$~km~s$^{-1}$ in core 6 to a maximum of $\sim 3.2$~km~s$^{-1}$ in core 2 (Table~\ref{tbl:core}). The three-dimensional velocity dispersion of the cores provided by the NH$_3$ line width lies in the range of 1.2 -- 2.4~km~s$^{-1}$ which is less than the escape velocity in each core. However, the C$^{18}$O velocity dispersions (2.5 - 4~km~s$^{-1}$) exceed the escape velocity in some regions of the cloud, especially for core 6, which has a small escape velocity. This may lead to instability for core 6. Additionally, an expansion speed of 1.0~km~s$^{-1}$ in the envelope of the core (cf. \S\ref{sec:lin}), though significantly less than the escape velocity, may aggravate the situation.

To examine the gravitational stability of the cores, it is useful to calculate the virial mass of the cores as well as the virial parameter, the ratio between the virial mass and core mass. The virial parameter given by \citet{ber92} can be expressed as $\alpha=5\sigma^2R/GM$, where R is the radius of the core, $\sigma=\sqrt{3/(8\ln2)}\times FWHM$, and $M$ is the core mass. The virial mass and virial parameter can be found in Table~\ref{tbl:core}. The average virial parameter is about 1.1 which suggests that most of the cores are virialised. However, the virial parameters of cores 2 and 6 deviate significantly from unity. This can be caused by a number of factors. One explanation is the potential for the presence of spatially unresolved stellar clusters in the cores, which can be identified from the color temperature of the dust continuum map. For $M_{\rm Virial}>M_{\rm Molecular}$, beam filling factors smaller than unity or streaming motions may adversely affect the estimate and may lead to an underestimated column density or an overestimated line width. Alternatively, it is possible that the cores may not actually be in virial equilibrium, and may be transient entities \citep{bal06}, which is probably the case in core 6.

We derive a Jeans mass greater than 170~$M_\sun$ for each core using the equation $M_{Jeans} = 17.3~{T_{kin}}^{1.5} n^{-0.5} M_\sun$, where $T_{kin}$ is the kinetic temperature derived from NH$_3$ data, and $n$ is the average density as listed in Table.~\ref{tbl:core}. We note that the masses of cores 1 and 2 are comparable to their Jeans masses, but the other cores are significantly less massive. However, considering the clumping in cores 1 and 2, it is still possible for them to form several individual protostars. Therefore, we suggest that cores 1 - 5 are gravitationally bound, among which cores 1 and 2 are marginally stable against collapse. However, core 6 is probably transient.

\subsection{Comparison with more evolved clouds}

Our dark cloud has a lower kinetic temperature than other evolved clouds. \citet{wu06} reported a mean temperature of 19~K in massive water maser sources excluding known HII regions, and in ultra-compact HII (UC~HII) regions. \citet{chu90} give a significantly higher temperature spread from 15~K to $>$60~K for UC~HII regions. The temperature of the cloud is a good tracer to determine its evolutionary stage in the context of massive star formation. The low temperatures in our cloud suggests that it is not an active high-mass star forming region yet.

The C$^{18}$O line widths are typically a factor of two to five times smaller than the line widths in more evolved massive star forming regions traced by other molecules with similar densities. The NH$_3$ line widths we find are comparable to the massive cores reported by \citet{pil06} and \citet{wu06}, but are smaller than those associated with water masers or are in proximity to UC~HII regions. More specifically, the line widths of our cloud are similar to those of the most quiescent NH$_3$ cores with no methanol maser or 24~GHz continuum emission described by \citet{lon07} and smaller than those in more evolved cores with maser or continuum emission. Our cloud also presents similar NH$_3$ line width to those bright-rimmed clouds which are undergoing recently initiated star formation and being subjected to intense levels of ionizing radiation \citep{mor10}. After removing the thermal broadening contributions to the line widths, the velocity dispersions associated with turbulence are supersonic and lie between those of the sources triggered by the radiatively driven implosion and the non-triggered ones. For the more evolved clouds in the UC~HII phase, \citet{chu90} reported significantly larger line widths of $\sim 3$~km~s$^{-1}$. The large line widths in more evolved stages are likely due to a combination of warmer temperatures and broadening from dynamics such as outflows.

The HCO$^+$~(1$-$0) spectra associated with hyper-compact H~II regions have a considerably larger line width (15-60~km~s$^{-1}$) reported by \citet{chu10}. For our cloud, our HCO$^+$ line profiles are similar to the samples of \citet{bra01} and \citet{ces99}, which are all in the pre-UC~HII stage of star formation, though some sources in \citet{ces99} are proven to be more evolved than those in \citet{bra01}. We find similar line widths and profiles as theirs, with a self-absorption dip between the blue and red peak. However, our cloud is colder than the pre-UC~HII sources and do not have any associated IRAS point source.

The optical depths of NH$_3$~(1,1) are comparable to the cores given by \citet{pil06} and \citet{lon07}. Our core 4, 5 and 6 have a lower optical depth than the NH$_3$ sources ($\sim 2.7 - 2.9$) associated with methanol masers or 24-GHz continuum emission in \citet{lon07}, while core 1, 2 and 3 near the cloud center gives higher optical depth but still lower than those samples ($\sim 3.1$) with only NH$_3$ association. However, due to the large uncertainty, without further observation, it would be premature to classify these sources into more detailed evolutionary stages based on optical depth. Our cloud density is about an order of magnitude lower than the typical value in more evolved massive star-forming regions: an average density of $10^6$~cm$^{-3}$ is given by \citet{beu02} using LVG calculations in clouds prior to building up an UC~HII region, and a similar value is given by \citet{pil07} in clouds harboring a UC~HII region.

Our NH$_3$~(1,1) core sizes lie between the mean value of 0.28~pc given by \citet{lon07} and 0.57~pc given by \citet{pil06}, but are significantly smaller than the mean size of 1.6~pc reported by \citet{wu06}. The small values of \citet{lon07} probably result from their small beam size (11\arcsec), but the large values of \citet{wu06} are likely a result of evolution. All these aspects suggest that our cloud is in an early evolutionary stage of massive star formation. These quiescent cores are likely the candidates for pre-protostellar cores.

\subsection{Rotation in the cores}\label{sec:rot}

One of the possible generators of the velocity gradients mentioned in \S\ref{sec:vel} is rotation in the cores. \citet{goo93} has discussed the relation between rotation and the geometric properties of the cores. They found no causal relation between velocity gradient direction and core elongation, nor any relationship between the magnitude of the gradient and core shape, on the size scale of $10^{17}$~cm. An examination of the gradients in our clouds also reveals no obvious relation of this kind. In view of the early evolutionary state and the dominant role of non-thermal line broadening in the cloud, we believe that the velocity gradients in the cloud are more affected by stochastic processes, such as fragmentation, collisions, and nonuniform magnetic fields rather than ordered motions such as rotation.

We use the parameter $\beta$ as defined by \citet{goo93} to compare the rotational kinetic energy to the gravitational energy. Thus $\beta$ can be written as
\begin{displaymath}
\beta={(1/2)I\omega^2 \over qGM^2/R}={1 \over 2}{p \over q}{\omega^2R^3 \over GM},
\end{displaymath}
where $I$ is the moment of inertia given by $I=pMR^2$, $qGM^2/R$ represents the gravitational potential energy of the mass $M$ within a radius $R$, and $\omega=\mathscr{G}/\sin i$, where $i$ is the inclination of the core along the line of sight. We assume $p/q=0.22$ as for a sphere with an $r^{-2}$ density profile and $\sin i=1$. Our $\beta$ values as listed in Table~\ref{tbl:vgradient} are consistent with the result of \citet{goo93} that most clouds have $\beta\leq 0.05$. The small values of $\beta$ show that the effect of rotation is not significant in maintaining the overall dynamical stability for the cloud. The results also indicate that these cores are unlikely to experience instabilities driven by rotation (e.g., bars, fission, or rings), if the magnetic fields are not taken into account.

\section{Summary}\label{sec:sum}
The multiple molecular line observations of the infrared dark cloud MSXDC G084.81$-$01.09 were analyzed. The results are summarized as below.

The cloud has a low gas temperature of 12~K, a column density of $10^{22}$~cm$^{-2}$ and masses of the cores range from 60 to 250~$M_\sun$. The spatial distribution of the ammonia emission correlates well with the mid-infrared absorption and millimeter dust emission. All 6 ammonia cores are associated with their corresponding dust peaks. The line width of 1~km~s$^{-1}$ for ammonia indicates the dominant role of non-thermal broadening in the cloud. The abundances of ammonia range from $2 - 3.5\times10^{-8}$. These facts show that the cloud is a site of potential massive star formation at a very early evolutionary stage.

$^{13}$CO and C$^{18}$O trace a more extended cloud structure. Together with the HCO$^+$ emission, we detect an expanding envelope in part of the cloud with a expansion velocity of 1~km~s$^{-1}$. Five of the cores are gravitationally bound (four are virialised) with Jeans masses comparable or exceeding their molecular mass except a possible unbound transient core. We found velocity gradients in the cores using different molecular tracers. The different directions and magnitudes of the gradients using the different tracers indicate clumping motions on different scales.

The observations presented here are part of a larger program designed to search   and investigate potential massive star forming regions. Higher resolution observations with interferometers would be needed to study turbulence and fragmentation inside our cores. Additional submillimeter mapping observations could derive the dust temperature and structure associated with the gas. We could also use radio polarization observation to study the relationship between magnetic field and cloud structure at the early phrases of massive star forming.

\acknowledgements{This work is made based on observations with the 100-m telescope of the MPIfR (Max-Planck-Institut f\"ur Radioastronomie) at Effelsberg and Delingha 13.7-m telescope of Purple Mountain Observatory. This work is also based on observations made with the Spitzer Space Telescope, which is operated by the Jet Propulsion Laboratory, California Institute of Technology under a contract with NASA. The authors appreciate all the staff members of the observatories for their help during the observations. This work was supported by the Chinese NSF through grants NSF 11073054, NSF 10733030, NSF 10703010 and NSF 10621303, and NBRPC (973 Program) under grant 2007CB815403.

\clearpage

\begin{figure}
\epsscale{0.8}
\plotone{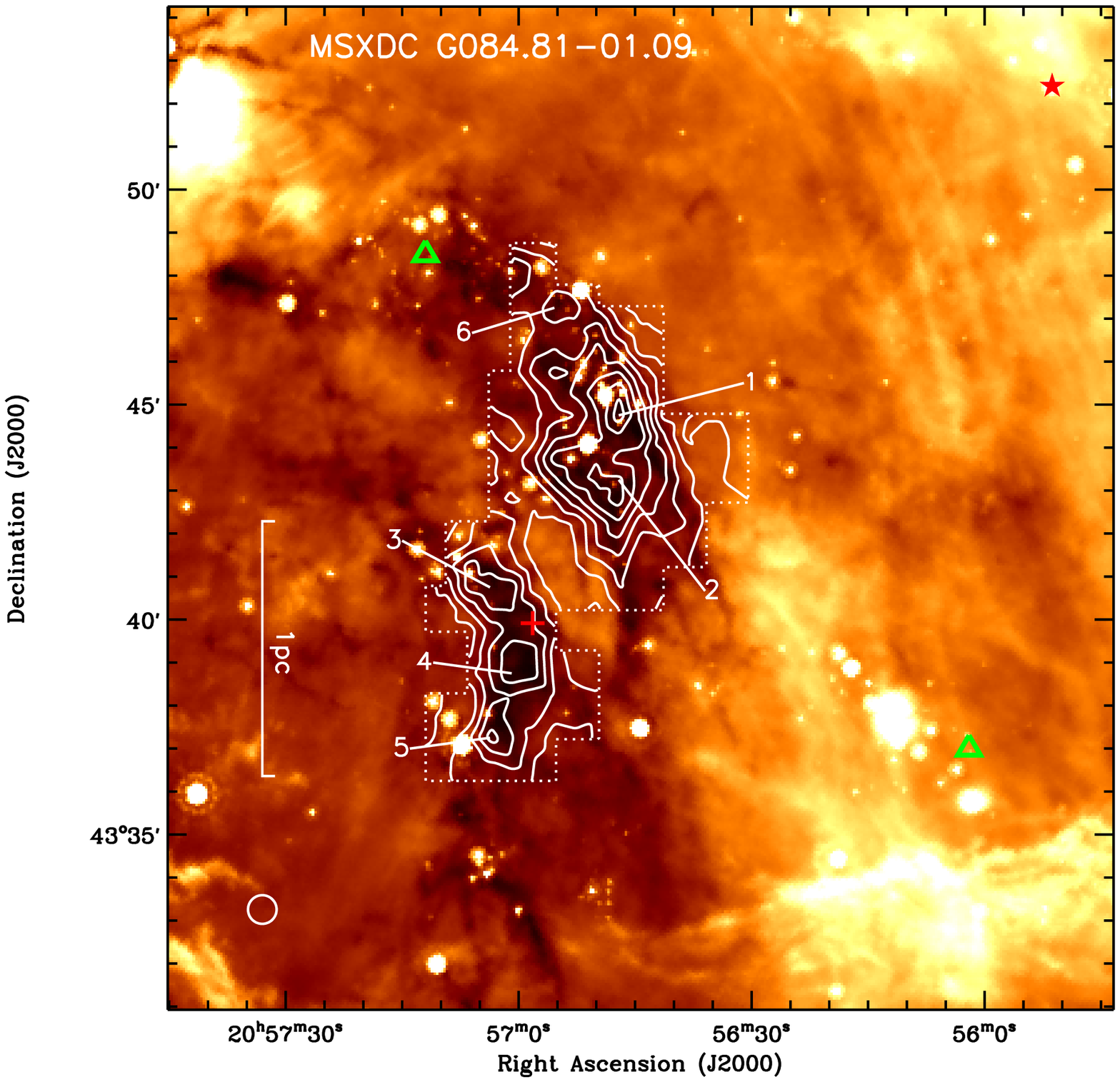}
\caption{Spitzer MIPS 24 $\mu$m image overlaid with an NH$_3$~(1,1) contour map of integrated intensity from the MSX dark cloud G084.81$-$01.09. The spectra were integrated over $-$2.5 to 4.5 km\,s$^{-1}$. The peak contour flux is 8 K\,km\,s$^{-1}$ with contour interval 1 K\,km\,s$^{-1}$. The red star indicates the ionizing star of W80 complex, 2MASS~J205551.25+435224.6, reported by \citet{com05}. The cross indicates the MYSO candidate, G084.7847$-$01.1709, identified by \citet{urq09}. The green triangles indicate the star clusters detected by \citet{cam02}. The white circle shows the beamsize of the NH$_3$~(1,1) observation.}
\label{fig:NH3+MIPS1}
\end{figure}

\begin{figure}
\epsscale{0.8}
\includegraphics[width=0.5\textwidth]{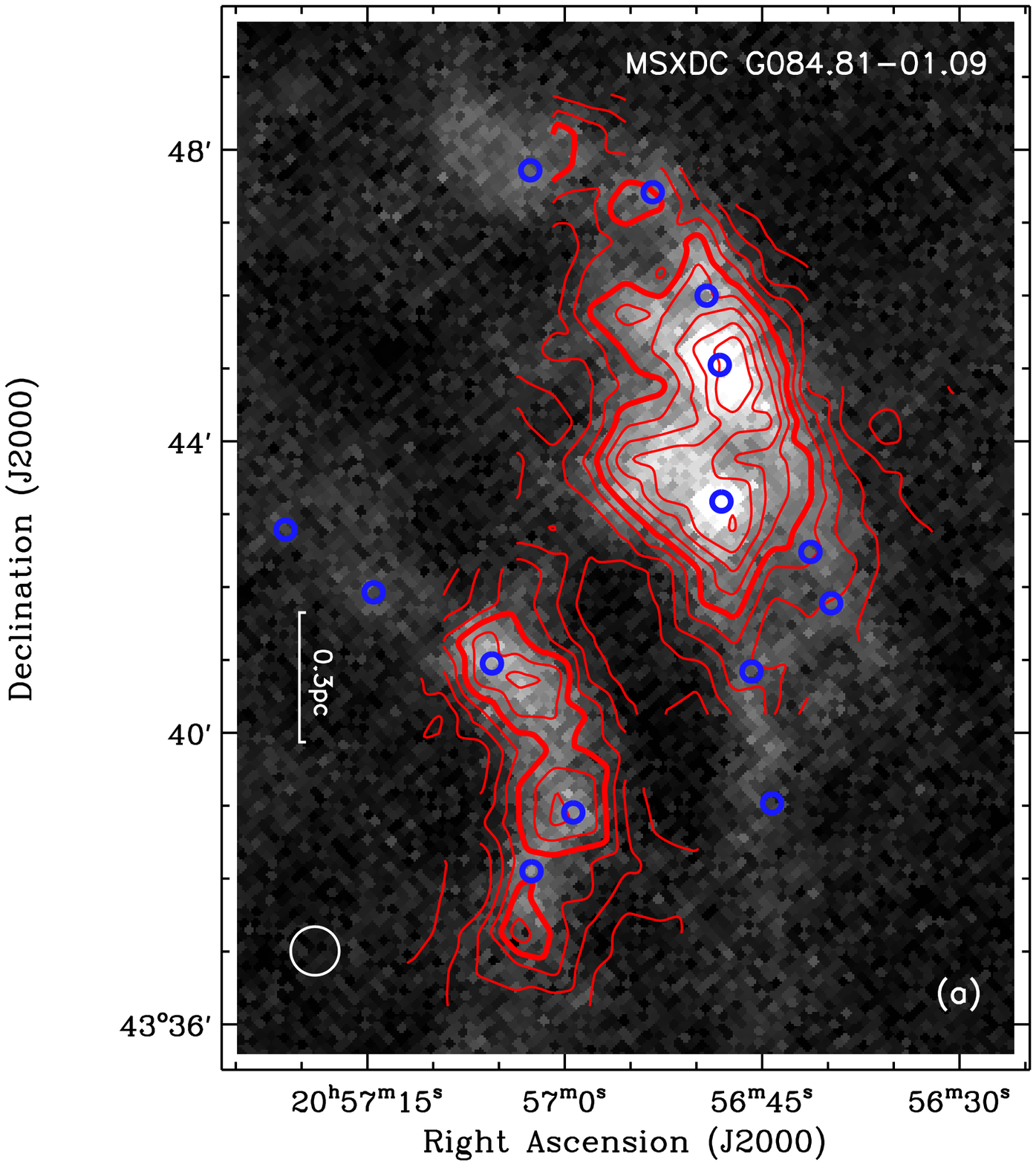}
\includegraphics[width=0.5\textwidth]{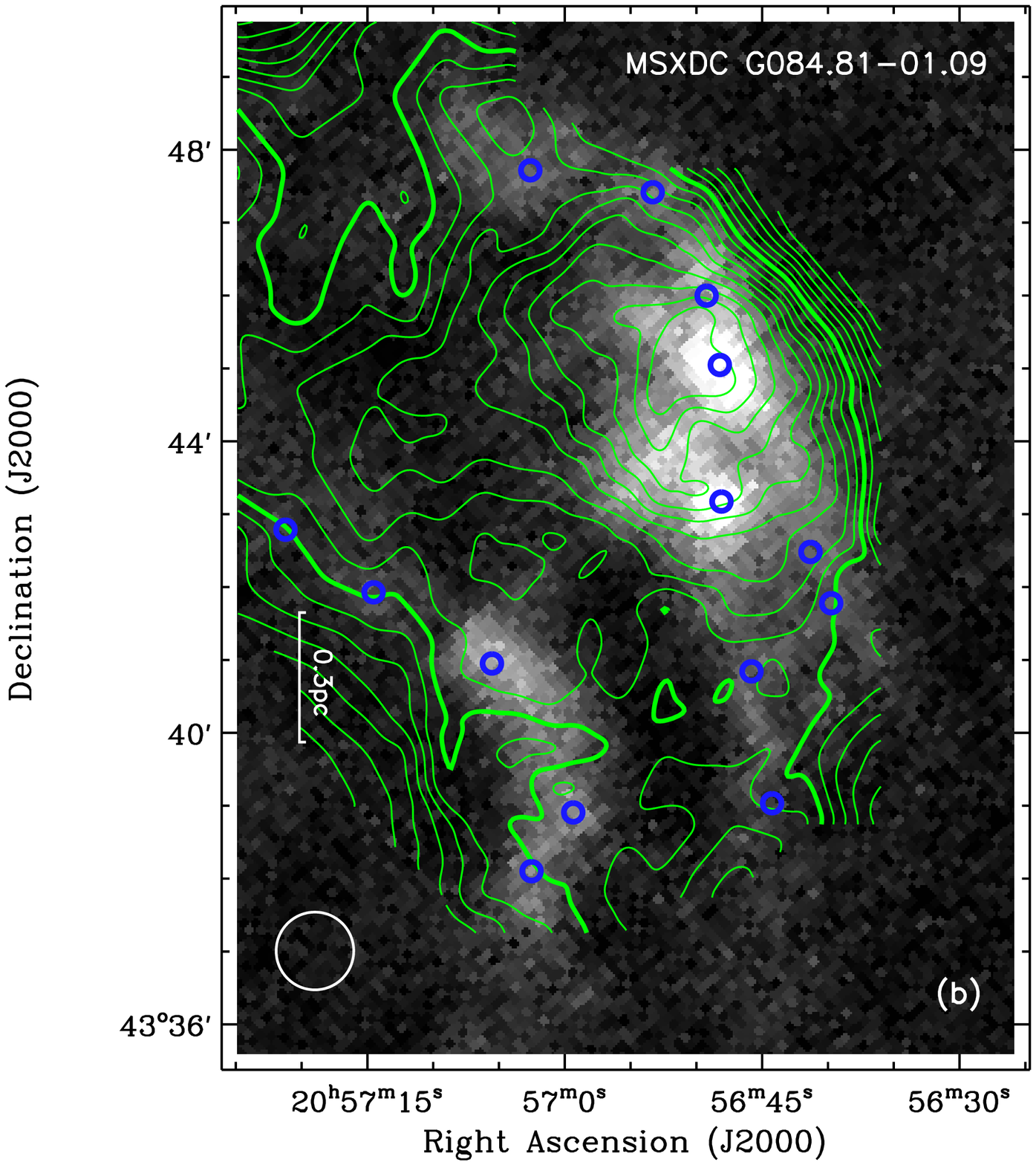}\\
\includegraphics[width=0.5\textwidth]{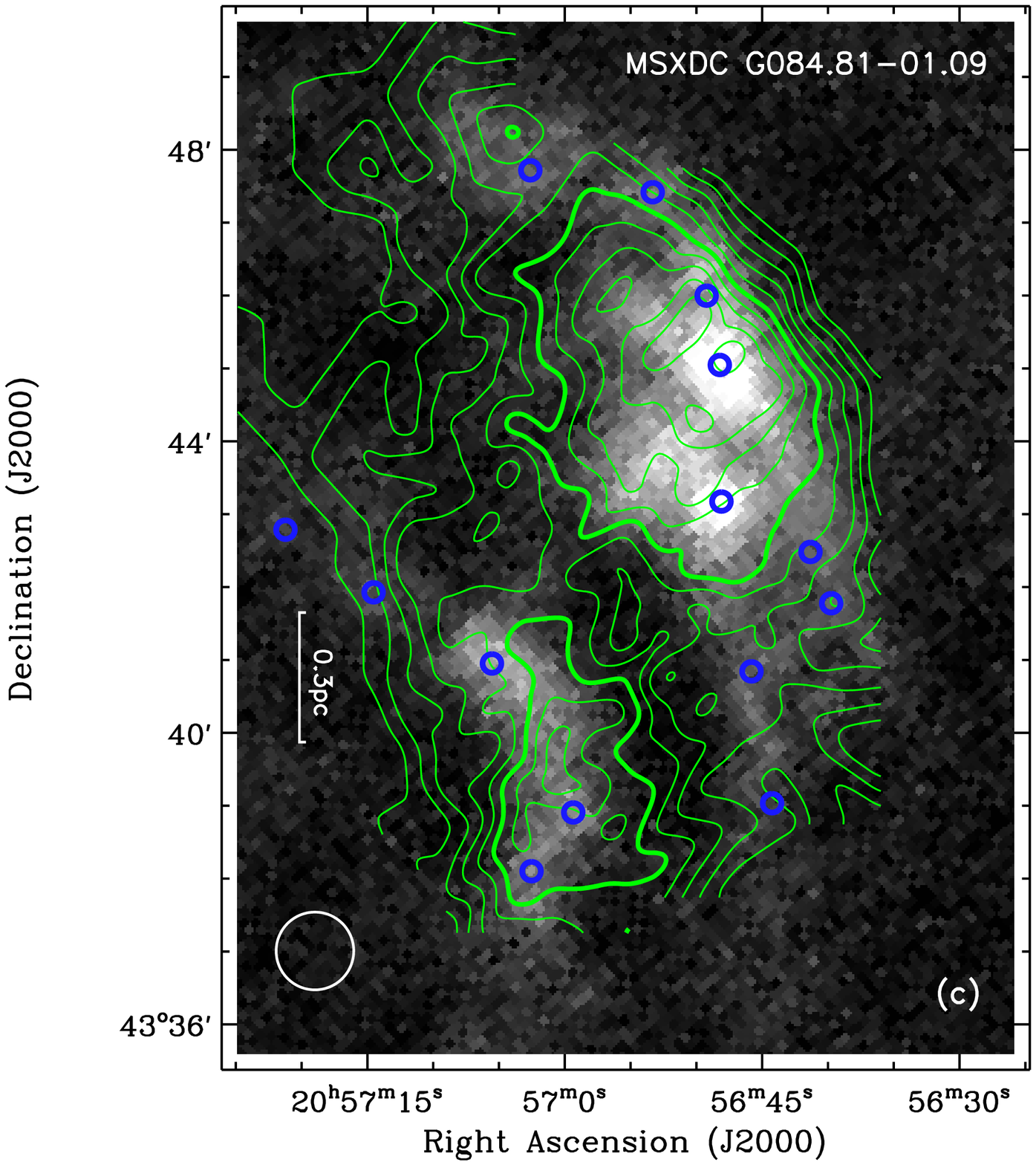}
\includegraphics[width=0.5\textwidth]{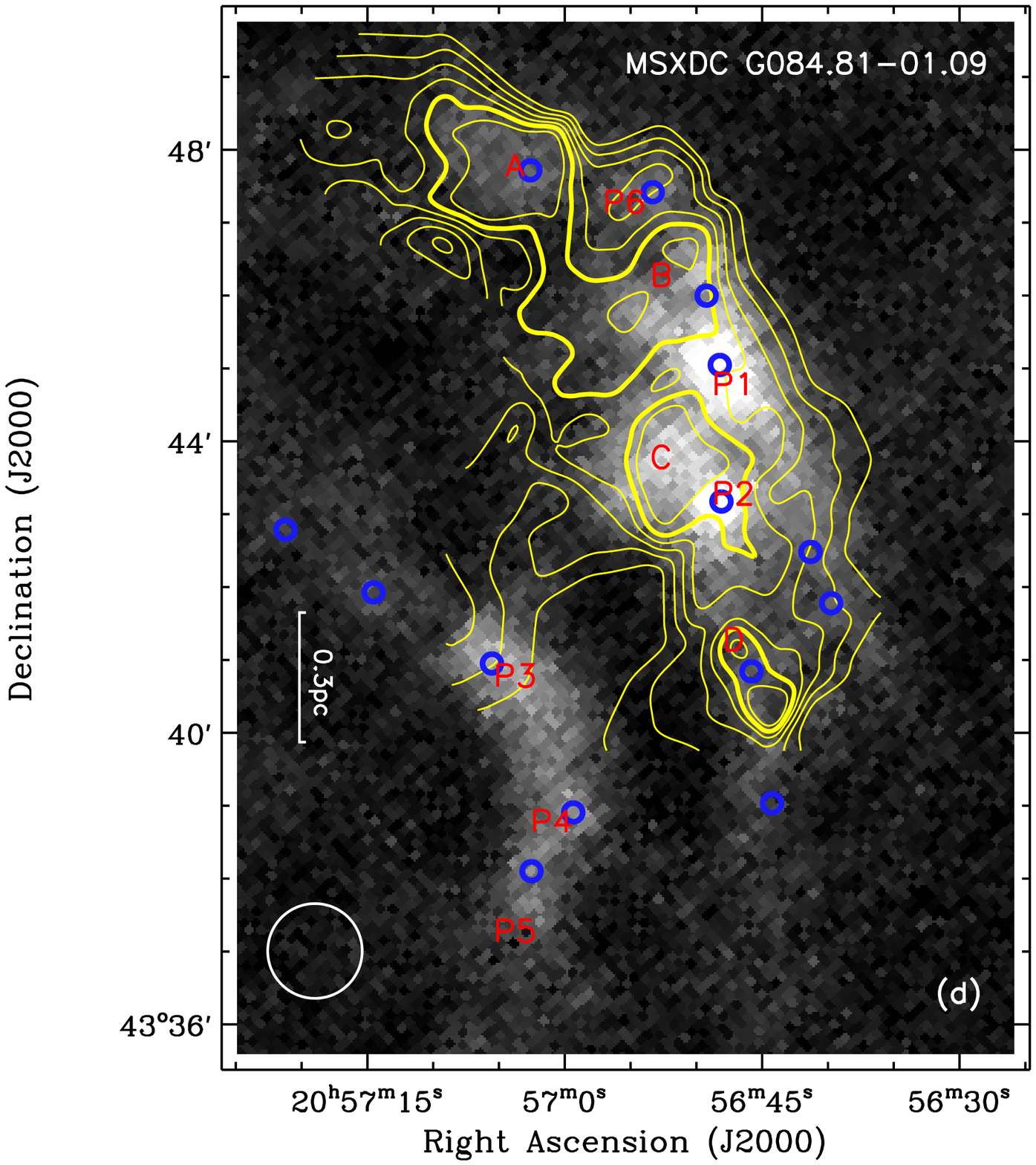}
\caption{CSO BOLOCAM 1.1 mm continuum image overlaid with an integrated intensity contour map of observed molecular lines. (a) NH$_3$~(1,1) integrated over $-$2.5 to 4.5 km\,s$^{-1}$, with a bold line at 50\% of maximum intensity of 8.7 K\,km\,s$^{-1}$ in steps of 9\%. (b) $^{13}$CO(1$-$0) integrated over $-$0.5 to 2.5 km\,s$^{-1}$, with a bold line at 70\% of maximum intensity of 38.2 K\,km\,s$^{-1}$ in steps of 5\%. (c) C$^{18}$O(1$-$0) integrated over $-$0.5 to 2.5 km\,s$^{-1}$, with bold line at 70\% of the maximum intensity of 8.3 K\,km\,s$^{-1}$ in steps of 6\%. (d) HCO$^+$(1$-$0) integrated over $-$2.5 to 4.5 km\,s$^{-1}$, with bold line at 70\% of the maximum intensity of 7.7 K\,km\,s$^{-1}$ in steps of 8\%. Four peaks in the HCO$^+$ map are designated as A-D. The cross indicates the MYSO candidate, G084.7847$-$01.1709, identified by \citet{urq09}. The small circles indicate the peak positions of clumps identified from the continuum map \citep{ros09}. The white circle at the bottom left shows the beamsize of the NH$_3$~(1,1) observation.}
\label{fig:MOL+CON}
\end{figure}

\begin{figure}
\epsscale{0.8}
\plotone{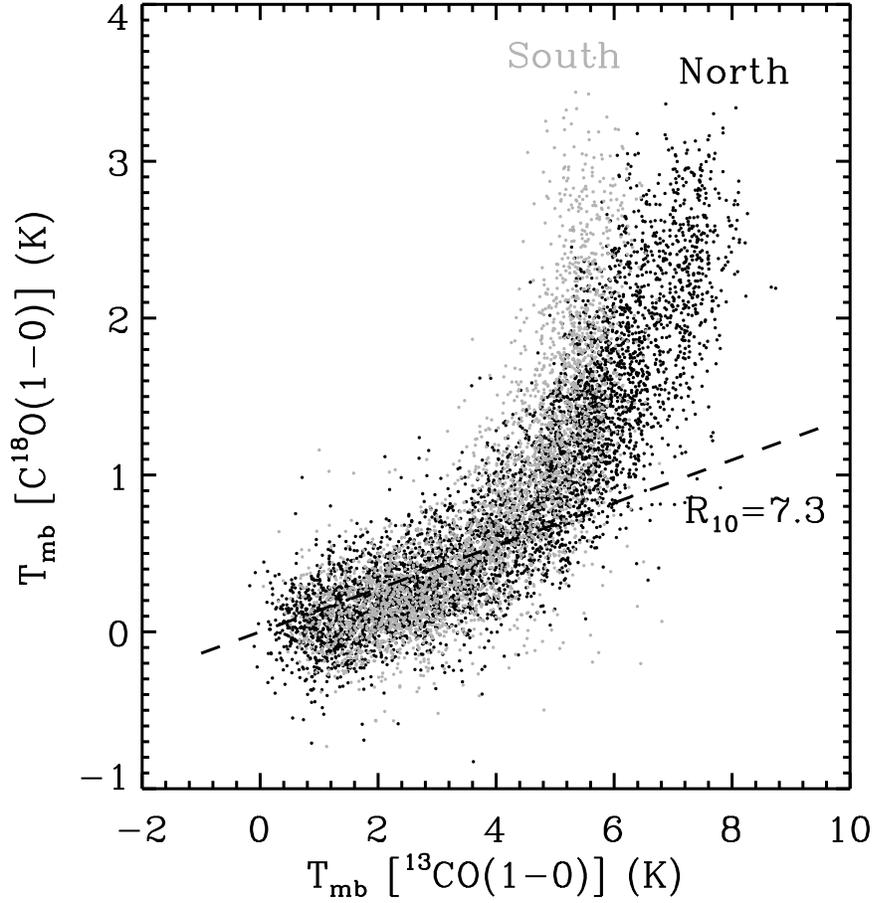}
\caption{$\rm T_{MB}$ of $^{13}$CO(1$-$0) vs. that of C$^{18}$O(1$-$0) for G084.81$-$01.09. The data are extracted within a velocity range of  $-$1.5 to 3.7~km~s$^{-1}$ and smoothed to 0.2~km~s$^{-1}$ spectral resolution. Each dot represents the $\rm T_{MB}$ of the same channel at same position for $^{13}$CO and C$^{18}$O. Data points from the northern and southern clumps are shown in black and grey colors respectively. The dashed-line represents the local ISM ratio. Note that the northern and southern clumps have a different trend for the intensity ratio at high intensities.}
\label{fig:TrCO}
\end{figure}

\begin{figure}
\epsscale{0.8}
\includegraphics[width=0.5\textwidth]{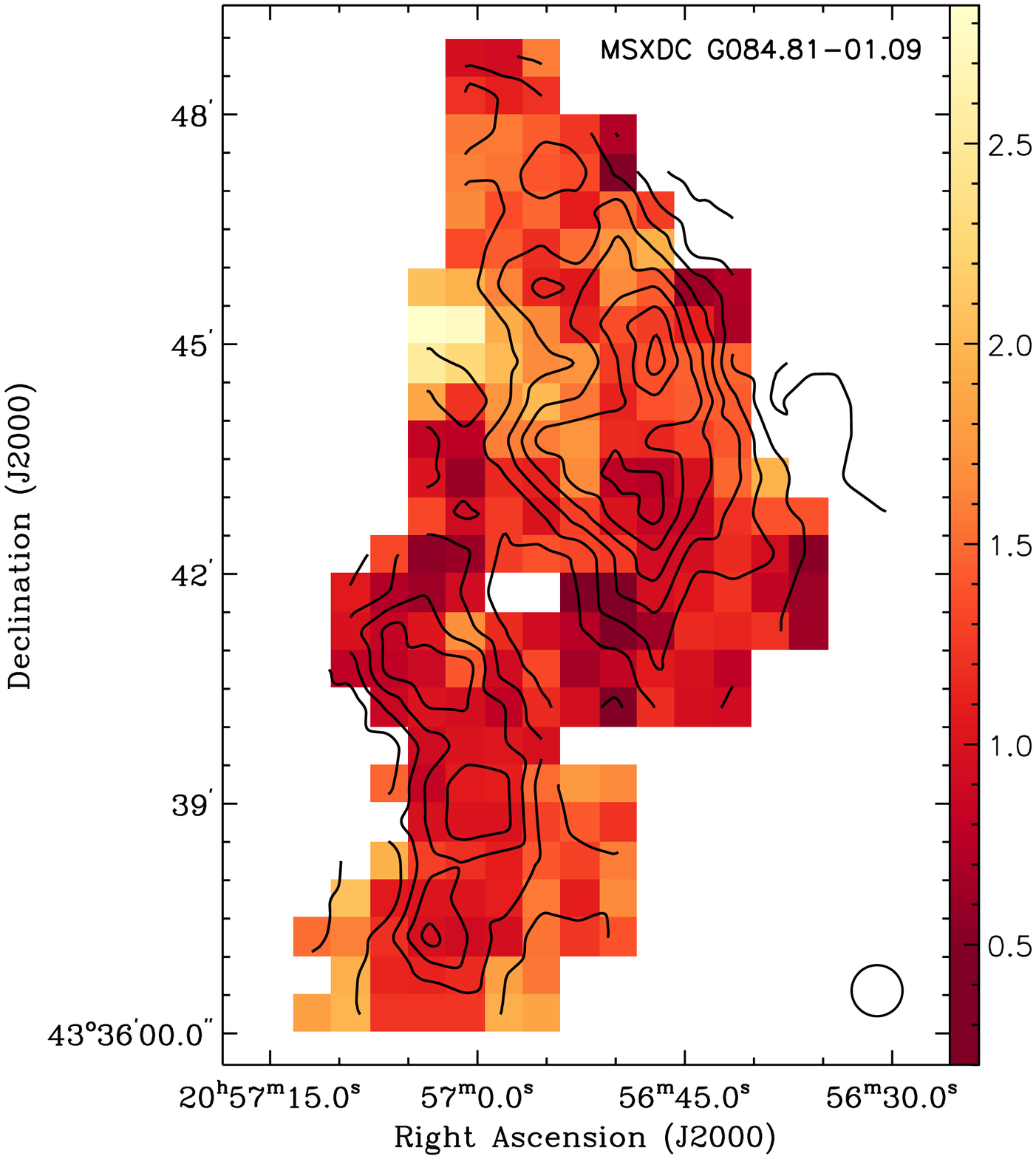}
\includegraphics[width=0.5\textwidth]{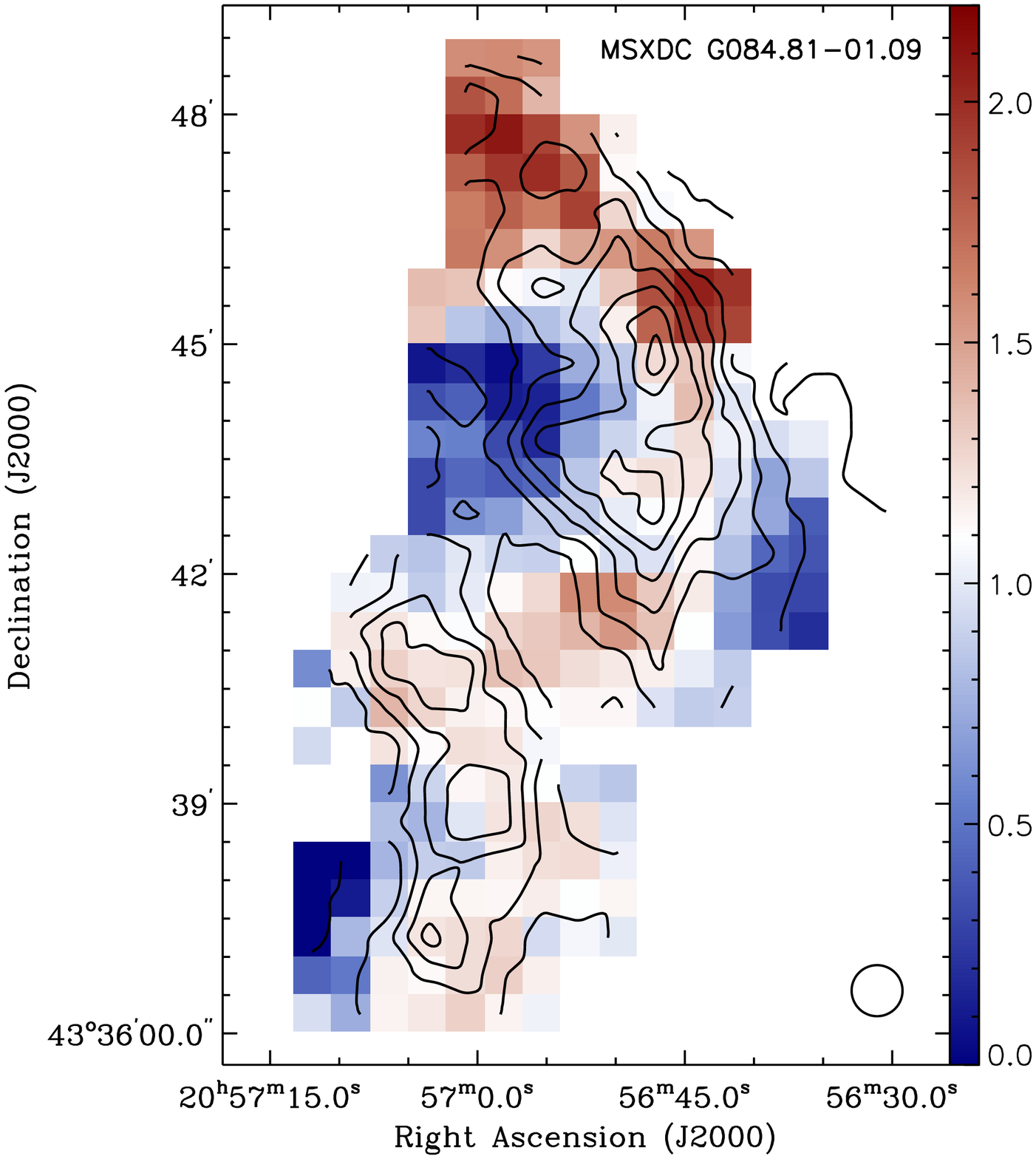}
\caption{Left panel: Color-coded map of the NH$_3$~(1,1) full width at half maximum (FWHM) line width in the MSX dark cloud G084.81$-$01.09 overlaid on a contour map of the integrated intensity. Right panel: Color-coded map of the local standard of rest velocity (V$_{\rm lsr}$) overlaid on a contour map of the integrated intensity. Only spectra with signal greater than 5$\sigma$ (FWHM map) and 3$\sigma$ (V$_{\rm lsr}$ map) are shown. The circle in the bottom right shows the beamsize.}
\label{fig:NH3 velocity}
\end{figure}

\begin{figure}
\epsscale{0.8}
\centering
\includegraphics[angle=270,scale=0.5]{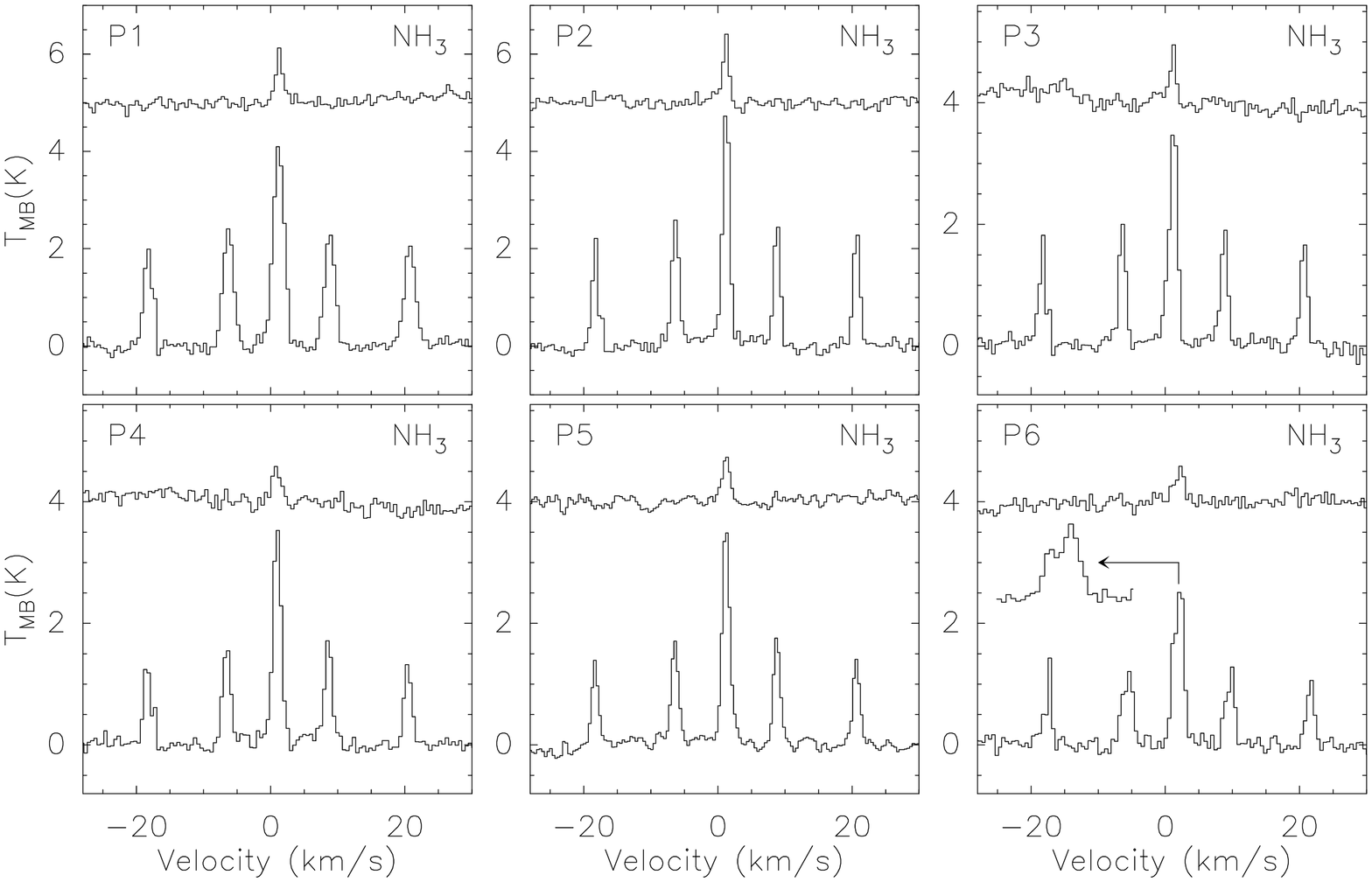}
\includegraphics[angle=270,scale=0.5]{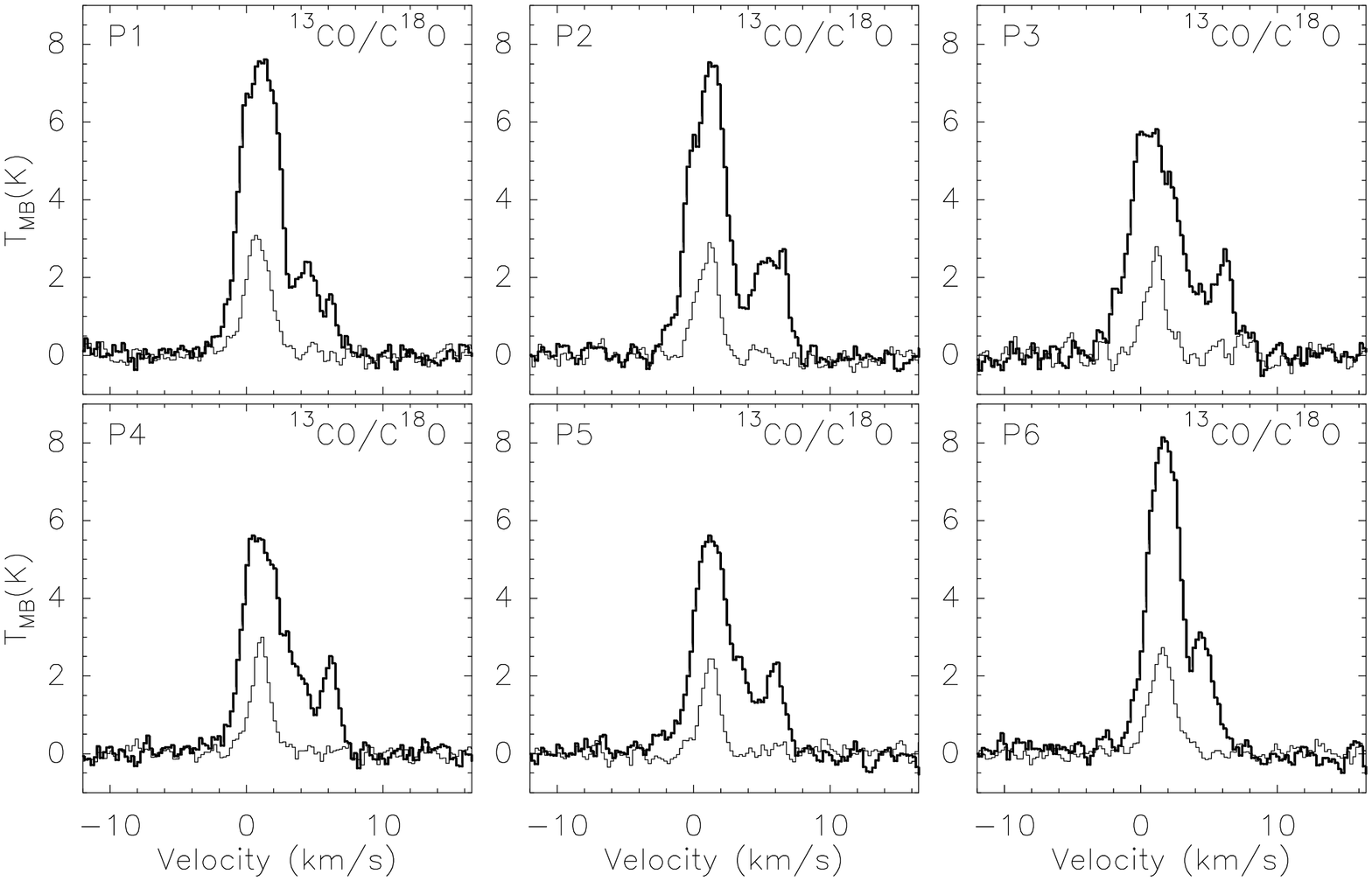}
\caption{Spectra of G084.81$-$01.09 toward the six peak positions (labeled as P1--P6). The first two rows show NH$_3$~(1,1) and (2,2) spectra. The NH$_3$~(2,2) lines are moved upward in each panel for clarity. The main (1,1) line feature of P6 was extracted and drawn beside the spectra. The next two rows show the $^{13}$CO (1$-$0) and C$^{18}$O (1$-$0) spectra. In all cases, the stronger line belongs to $^{13}$CO. The peak positions are referred to in Table~\ref{tbl:nh3 result}.}
\label{fig:spectra}
\end{figure}

\begin{figure}
\epsscale{0.8}
\centering
\includegraphics[scale=0.65,origin=br,angle=270]{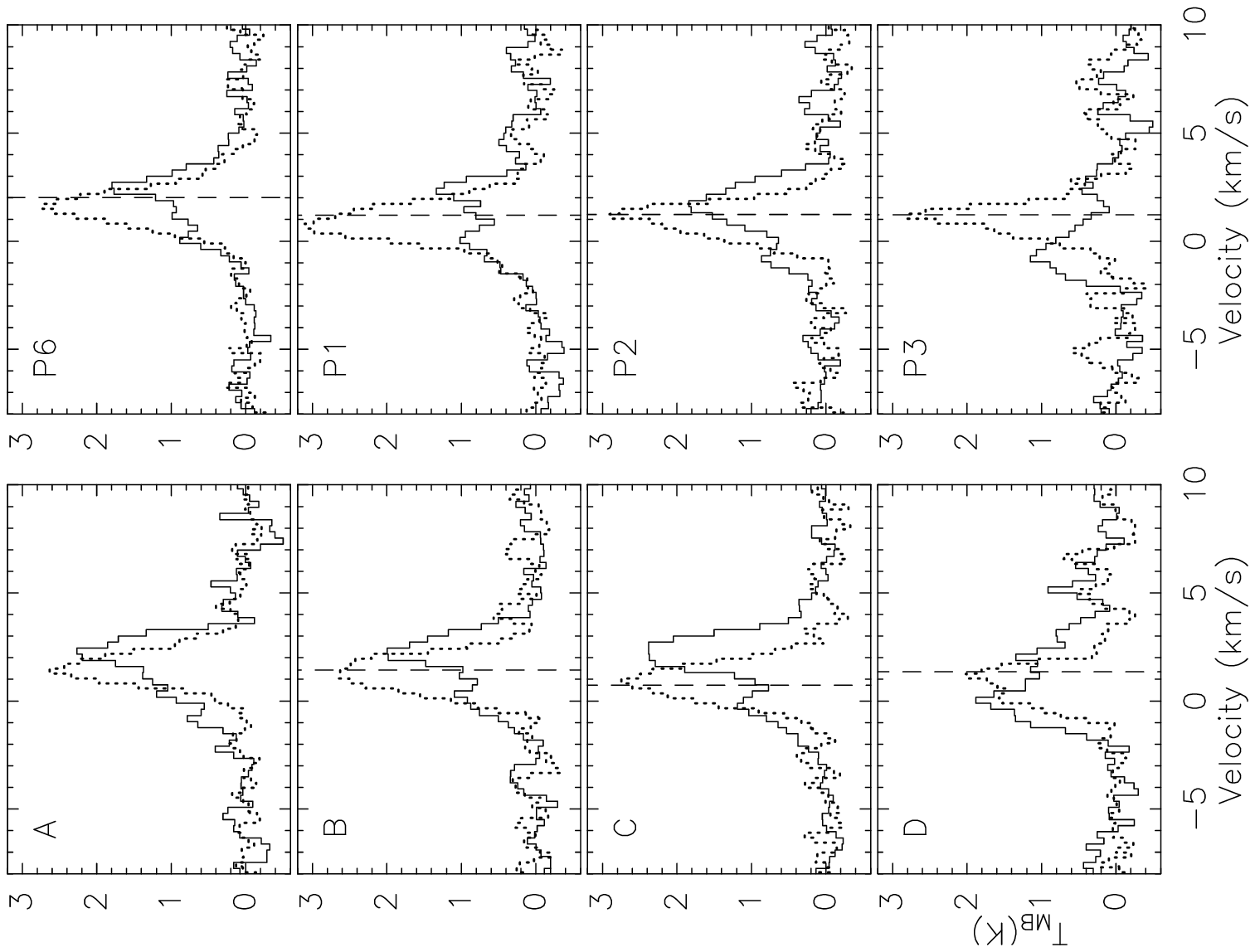}\ %
\includegraphics[width=0.52\textwidth]{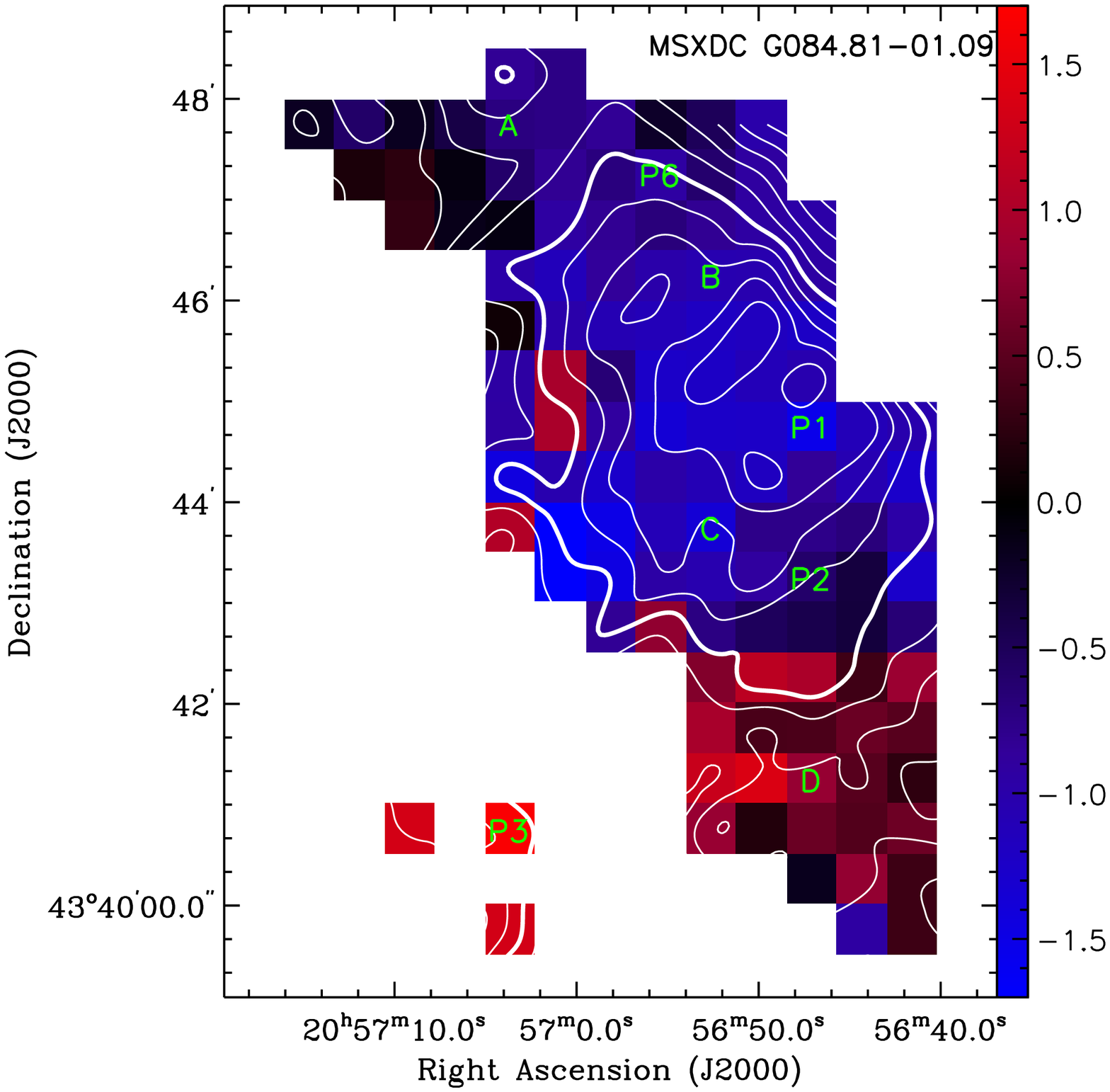} %
\caption{Left: HCO$^+$ (solid) and C$^{18}$O (dashed) spectra at the positions of the 4 NH$_3$ cores and the 4 HCO$^+$ peaks (indicated by red characters in Fig.~\ref{fig:MOL+CON}) from north to south. The vertical dashed lines indicate the V$_{lsr}$ of the NH$_3$~(1,1) line, obtained from a Gaussian fit. Right: Map of $\delta V$, the velocity difference between the peak of C$^{18}$O and HCO$^+$ overlaid on a C$^{18}$O contour map of the region.}
\label{fig:HCOP+C18O}
\end{figure}

\begin{figure}
\epsscale{0.8}
\centering
\includegraphics[angle=270,width=0.95\textwidth]{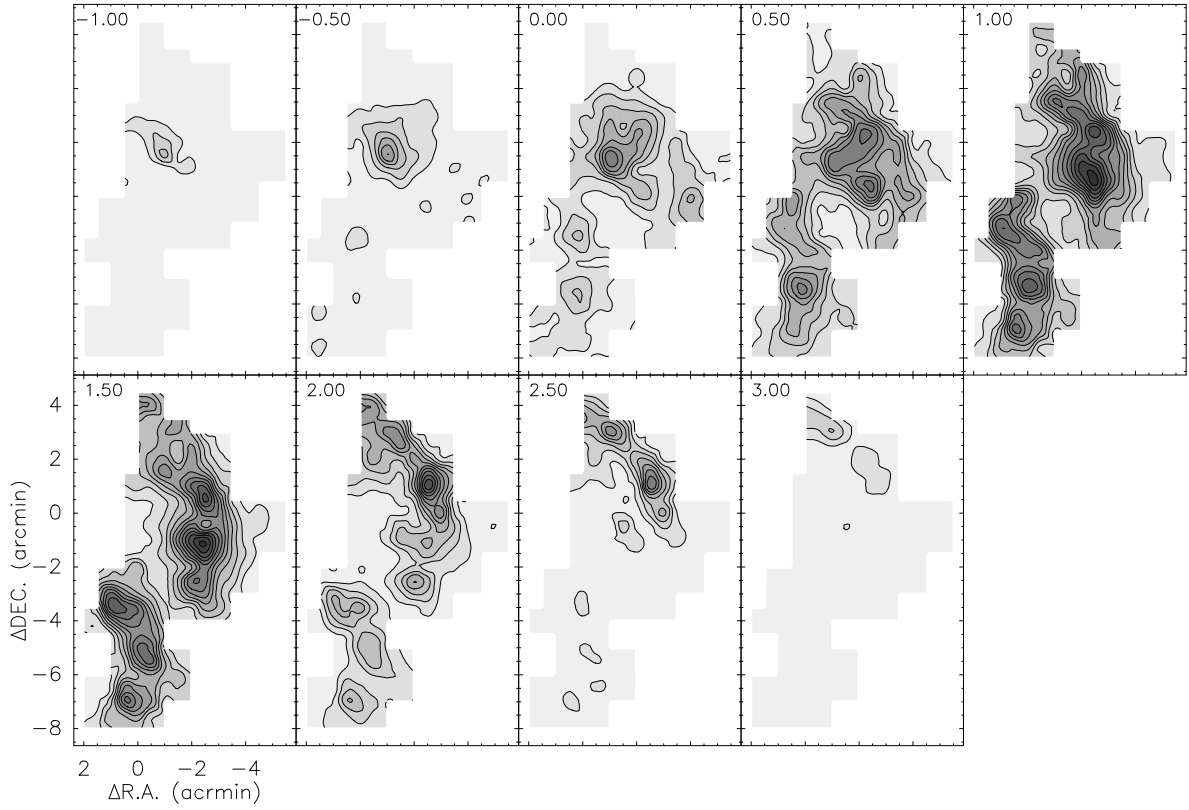}
\caption{NH$_3$~(1,1) channel maps for G084.81$-$01.09. The central velocity of each channel, in km\,s$^{-1}$, is marked on the top-left corner for each map. The reference position is at 20$^{\rm h}$57$^{\rm m}$00.9$^{\rm s}$,+43\arcdeg44\arcmin15.4\arcsec.}
\label{fig:NH3 channel}
\end{figure}

\begin{figure}
\epsscale{0.8}
\centering
\includegraphics[angle=270,width=0.80\textwidth]{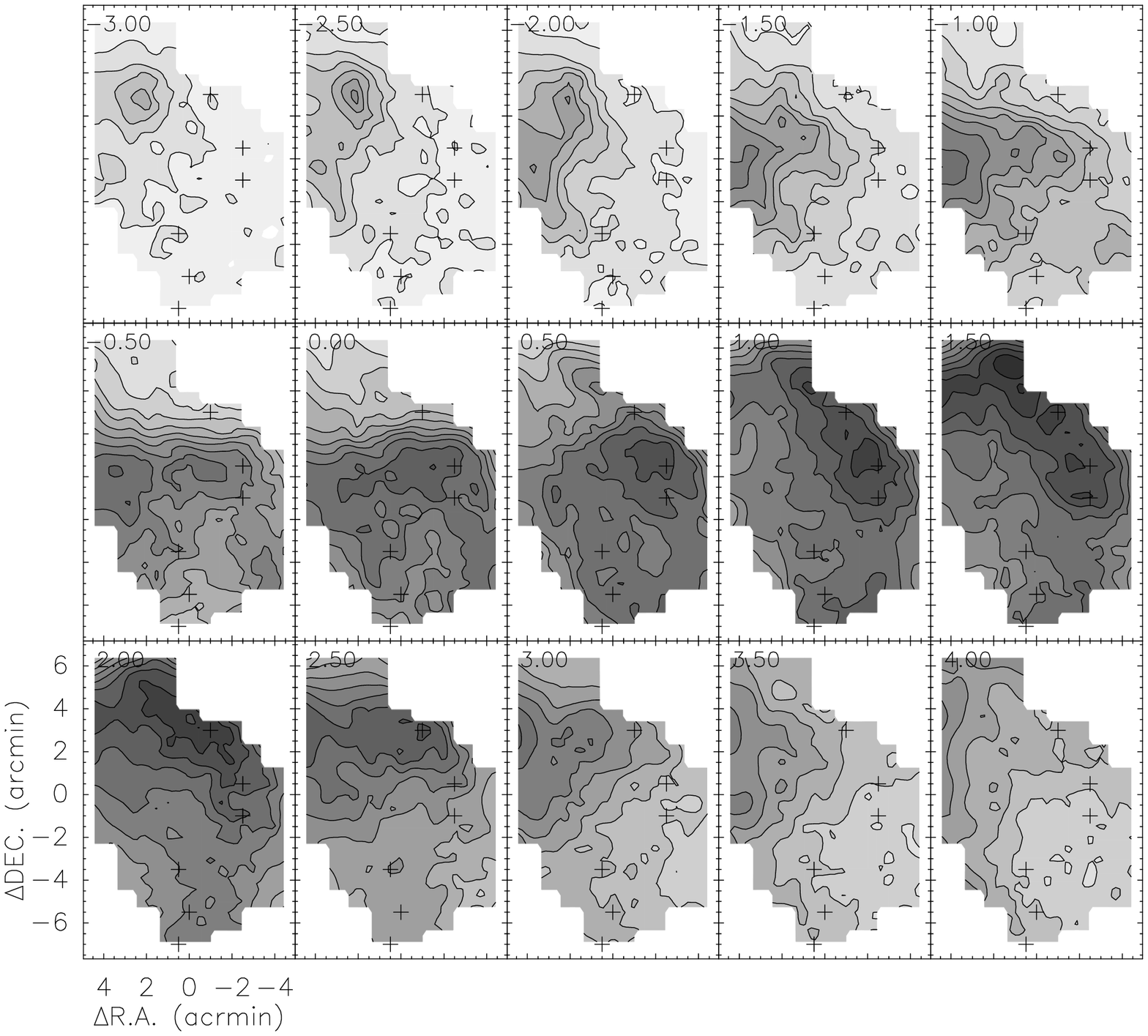}
\includegraphics[angle=270,width=0.96\textwidth]{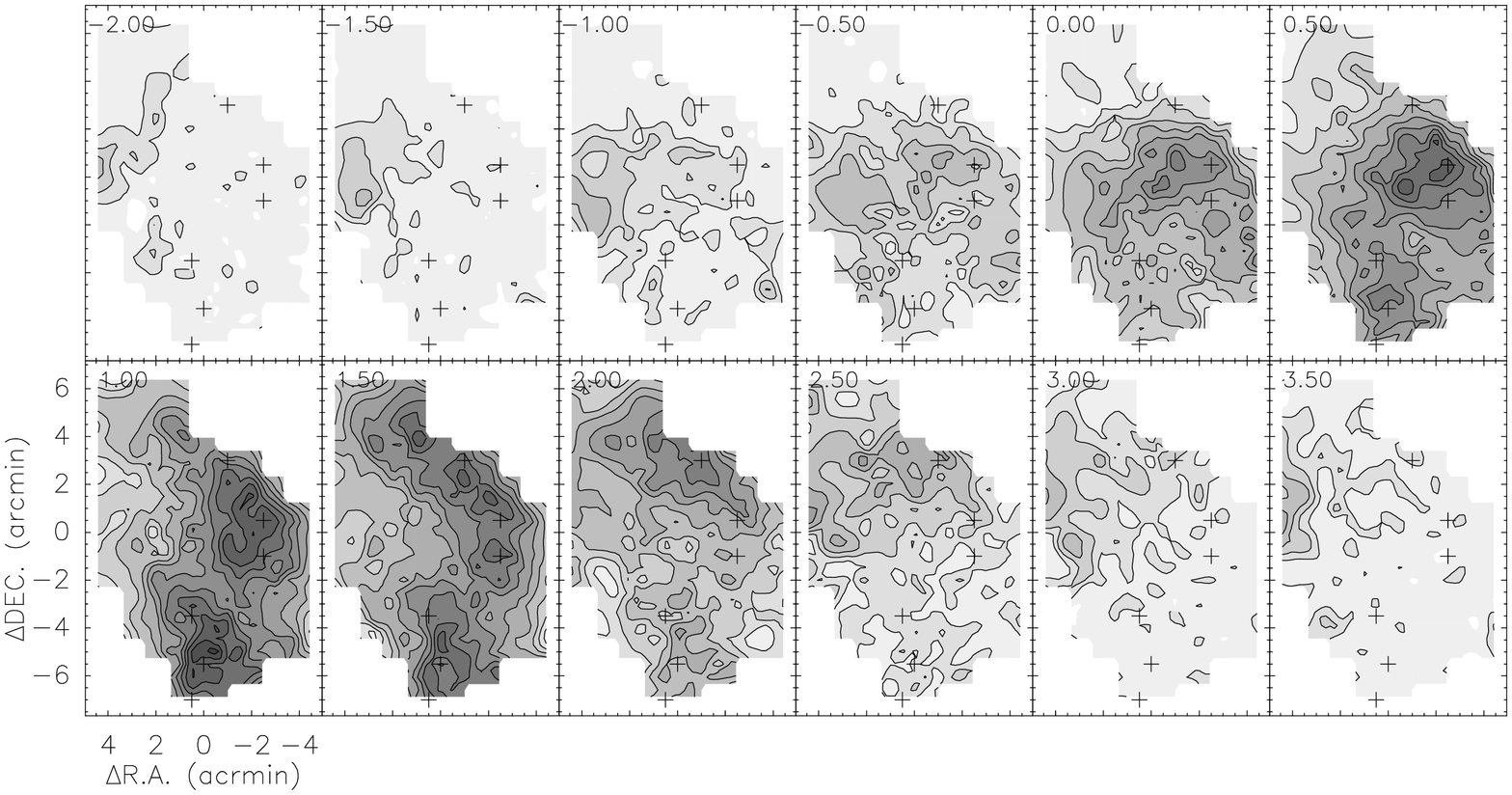}
\caption{$^{13}$CO(1$-$0) channel maps (upper panels) for G084.81$-$01.09 and C$^{18}$O(1$-$0) channel maps (lower panels) for G084.81$-$01.09. The central velocity of each channel, in km\,s$^{-1}$, is marked on the top-left corner for each map, and the plus signs indicate the positions of 6 cores identified from NH$_3$~(1,1) map (see Fig.~\ref{fig:NH3+MIPS1}). The reference position is the same as in Fig.~\ref{fig:NH3 channel}.}
\label{fig:CO channel}
\end{figure}

\begin{figure}
\epsscale{0.8}
\centering
\includegraphics[angle=270,width=0.85\textwidth]{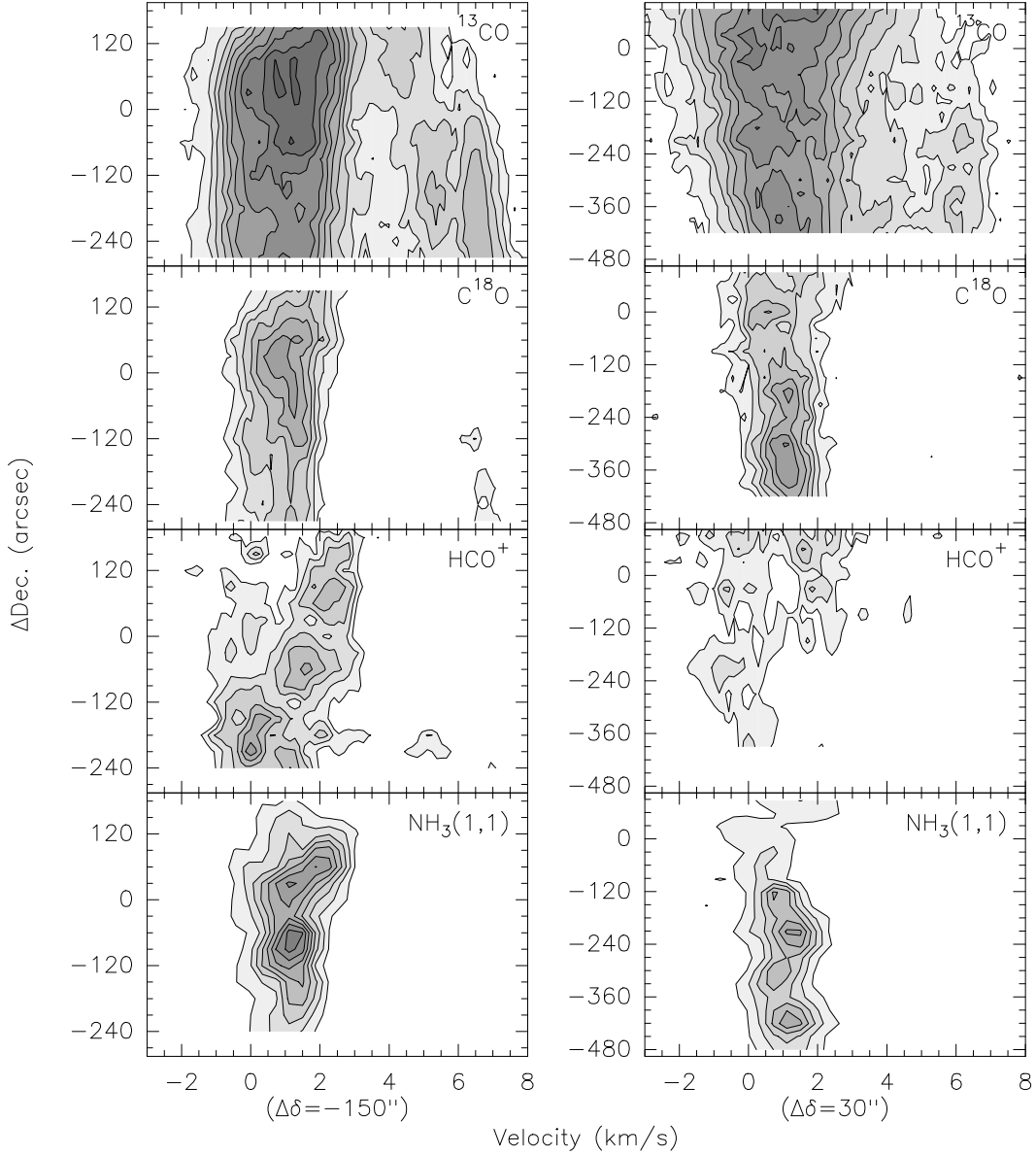}
\caption{The position-velocity map of $^{13}$CO (top), C$^{18}$O (second row), HCO$^+$ (third row) and NH$_3$~(1,1) main component (bottom) along the Declination axis for the MSX dark cloud G084.81$-$01.09, at Right Ascension of 20$^{\rm h}$56$^{\rm m}$47.1$^{\rm s}$ (left panel with offset $\Delta\delta$=$-$150\arcsec) and Right Ascension of 20$^{\rm h}$57$^{\rm m}$03.7$^{\rm s}$ (right panel with offset $\Delta\delta$=30\arcsec). The declination reference coordinate is the same as in Fig.~\ref{fig:NH3 channel}.}
\label{fig:pv map}
\end{figure}

\begin{figure}
\epsscale{0.8}
\plotone{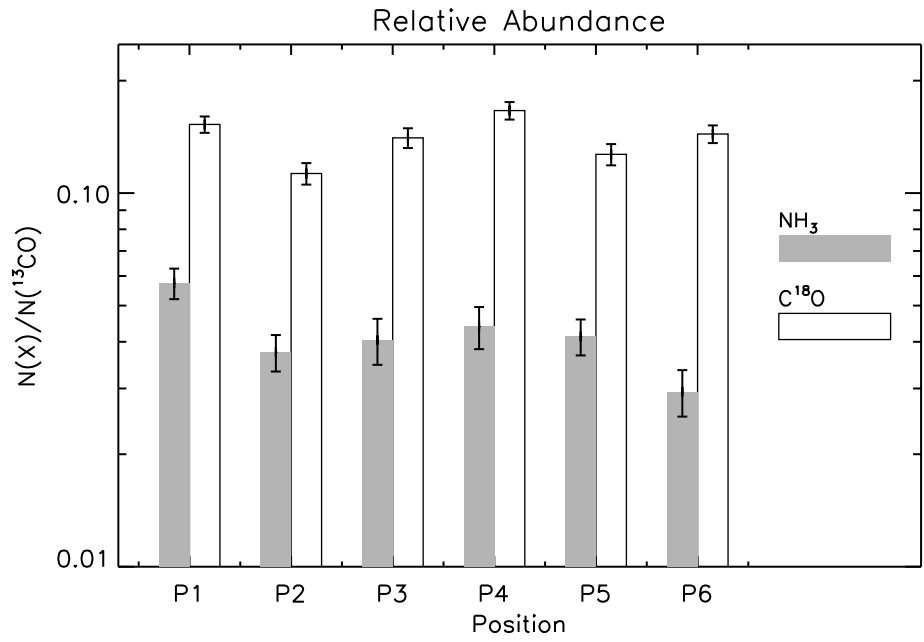}
\caption{Abundances relative to $^{13}$CO toward peak positions (Table~\ref{tbl:nh3 result}). The relative abundances were shown in different grey scale.}
\label{fig:relative abundance}
\end{figure}

\clearpage

\begin{figure}
\epsscale{0.8}
\plotone{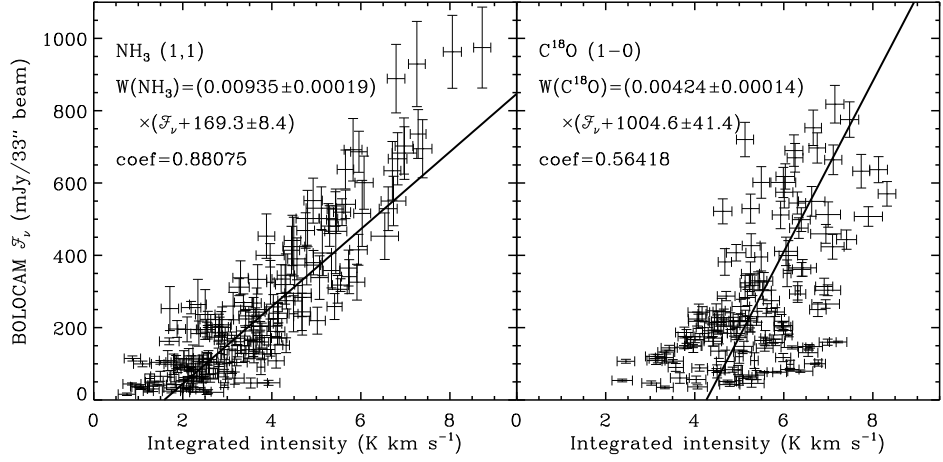}
\caption{Relation between the NH$_3$~(1,1) and C$^{18}$O~(1-0) integrated intensity and the 1.1~mm flux density. Continuum data were smoothed to corresponding resolution. The uncertanties of integrated intensity are the rms noise of the spectra and the error in dust flux is calculated following \citet{ros09}. The solid line represents the least-square fit of the data weighted in both coordinates.}
\label{fig:molevsdust}
\end{figure}

\clearpage

\begin{deluxetable}{lrllcrcr}
\tabletypesize{\scriptsize}
\tablecolumns{9}
\tablewidth{0pc}
\tablecaption{Observation parameters\label{tbl:observation}}
\tablehead{
\colhead{Line} & \colhead{$\nu_0$} &
\colhead{Telescope} & \colhead{During}  &
\colhead{HPBW} &
\colhead{$\delta \nu$} &
\colhead{$\delta v$} &
\colhead{$T_{\rm sys}$}
\\
\colhead{} & \colhead{(GHz)} &
\colhead{} & \colhead{} &
\colhead{(\arcsec)} &
\colhead{(KHz)} &
\colhead{(km~s$^{-1}$)} &
\colhead{(K)}
}
\startdata
NH$_3$ ($J, K$=1, 1) &  23.694496 & Effelsberg 100m & 2007 Dec. & 40 & 19.5 & 0.24 & 30-150 \\
                  &            &                 & 2008 Feb. & 40 & 30.5 & 0.38 & 30-45 \\
NH$_3$ ($J, K$=2, 2) &  23.722633 & Effelsberg 100m & 2007 Dec. & 40 & 19.5 & 0.24 & 30-140 \\
                  &            &                 & 2008 Feb. & 40 & 30.5 & 0.38 & 30-45 \\
NH$_3$ ($J, K$=3, 3) &  23.870130 & Effelsberg 100m & 2007 Dec. & 40 & 19.5 & 0.24 & 30-140 \\
                  &            &                 & 2008 Feb. & 40 & 30.5 & 0.38 & 30-45 \\
NH$_3$ ($J, K$=4, 4) &  24.139417 & Effelsberg 100m & 2007 Dec. & 40 & 19.5 & 0.24 & 30-160 \\
$^{12}$CO ($J$=1-0) & 115.271204 & DLH 13.7m & 2008 Mar.-Apr. & 60 & 142.1 & 0.37 & 200-400 \\
$^{13}$CO ($J$=1-0) & 110.201353 & DLH 13.7m & 2008 Mar.-Apr. & 64 &  41.7 & 0.11 & 200-400 \\
C$^{18}$O ($J$=1-0) & 109.782183 & DLH 13.7m & 2008 Mar.-Apr. & 64 &  42.1 & 0.11 & 200-400 \\
HCO$^+$ ($J$=1-0)   &  89.188530 & DLH 13.7m & 2008 May.-Jun. & $\sim$78 &  42.1 & 0.14 & 270-420 \\
\enddata
\tablecomments{$\nu_0$ is the rest frequency of the line, HPBW is the half power beam width of the telescope, $\delta \nu$ and $\delta v$ represent the frequency and velocity resolutions}
\end{deluxetable}

\clearpage

\begin{deluxetable}{lcccrrrrr}
\tabletypesize{\scriptsize}
\tablecolumns{9}
\tablewidth{0pc}
\tablecaption{Cores detected in the NH$_3$ observations of G084.81$-$01.09
\label{tbl:nh3 result}}
\tablehead{
\colhead{Peak} & \colhead{$\alpha$} & \colhead{$\delta$} &
\colhead{Transition} & \colhead{$V_{\rm LSR}$} & \colhead{$T_{\rm MB}$} &
\colhead{$FWHM$} & \colhead{$\tau_{\rm main}$} & \colhead{$\int{T_A^* d v}$}
\\
\colhead{} & \colhead{(J2000)} & \colhead{(J2000)} &
\colhead{} & \colhead{(km\,s$^{-1}$)} & \colhead{(K)} &
\colhead{(km\,s$^{-1}$)} & \colhead{} & \colhead{(K\,km\,s$^{-1}$)}
}
\startdata
1 ...... & 20 56 47.1 & +43 44 45 & 1-1 &  1.25$\pm$0.01 &  3.84$\pm$0.24 & 1.38$\pm$0.03 & 3.63$\pm$0.20 &  8.73$\pm$0.20 \\
         &            &           & 2-2 &  1.30$\pm$0.04 &  1.14$\pm$0.19 & 0.96$\pm$0.14 &               &  1.39$\pm$0.19 \\
2 ...... & 20 56 47.1 & +43 43 15 & 1-1 &  1.23$\pm$0.01 &  4.98$\pm$0.37 & 0.75$\pm$0.02 & 3.61$\pm$0.24 &  7.26$\pm$0.19 \\
         &            &           & 2-2 &  1.20$\pm$0.03 &  1.51$\pm$0.17 & 0.88$\pm$0.09 &               &  1.61$\pm$0.19 \\
3 ...... & 20 57 03.7 & +43 40 45 & 1-1 &  1.22$\pm$0.01 &  3.67$\pm$0.36 & 0.89$\pm$0.04 & 3.25$\pm$0.29 &  6.03$\pm$0.19 \\
         &            &           & 2-2 &  1.20$\pm$0.04 &  1.07$\pm$0.19 & 0.63$\pm$0.08 &               &  1.09$\pm$0.19 \\
4 ...... & 20 57 00.9 & +43 38 45 & 1-1 &  0.99$\pm$0.01 &  3.71$\pm$0.31 & 1.00$\pm$0.03 & 2.22$\pm$0.20 &  5.91$\pm$0.18 \\
         &            &           & 2-2 &  0.84$\pm$0.10 &  0.58$\pm$0.14 & 1.40$\pm$0.28 &               &  0.78$\pm$0.18 \\
5 ...... & 20 57 03.7 & +43 37 15 & 1-1 &  1.20$\pm$0.01 &  3.52$\pm$0.25 & 0.97$\pm$0.03 & 2.32$\pm$0.17 &  5.53$\pm$0.10\tablenotemark{a} \\
         &            &           & 2-2 &  1.21$\pm$0.06 &  0.72$\pm$0.09 & 1.32$\pm$0.21 &               &  1.02$\pm$0.11 \\
6 ...... & 20 56 55.4 & +43 47 15 & 1-1 &  2.01$\pm$0.02 &  2.51$\pm$0.23 & 1.40$\pm$0.04 & 2.08$\pm$0.21 &  5.02$\pm$0.18 \\
         &            &           & 2-2 &  2.07$\pm$0.14 &  0.53$\pm$0.12 & 1.98$\pm$0.34 &               &  1.39$\pm$0.19 \\
\enddata
\tablecomments{Columns are peak number, offset position from map center, NH$_3$~(J,K) transition, the local standard rest velocity, main beam brightness temperature, full width at half maximum, optical depth and integrated intensity of the main line. Values in column (4), (5), (6) and (7) are the HFS or GAUSS fitting results and errors estimated in CLASS. Integrated intensities are calculated from $-$2.5 to 4.5 km\,s$^{-1}$ with errors derived from rms.} \tablenotetext{a}{Peak 5 was observed in FFT mode with lower noise compared to Peak 1-4 and 6 in frequency switch mode.}
\end{deluxetable}

\clearpage

\begin{deluxetable}{lrccccc}
\tabletypesize{\scriptsize}
\tablecolumns{7}
\tablewidth{0pc}
\tablecaption{Physical properties of the NH$_3$ cores\label{tbl:nh3 property}}
\tablehead{
\colhead{Number}           &
\colhead{$T_{\rm ex}$}     &
\colhead{$T_{\rm rot}$\tablenotemark{a}}    &
\colhead{$T_{\rm kin}$\tablenotemark{b}}    &
\colhead{$N(\rm NH_3)$\tablenotemark{c}}    &
\colhead{$\Delta v_1$\tablenotemark{d}}     & \colhead{$\Delta v_2$\tablenotemark{e}}
\\
\colhead{}                        & \colhead{(K)}                       &
\colhead{(K)}                     & \colhead{(K)}                       &
\colhead{($10^{14}\rm cm^{-2}$)}  &
\colhead{(km\,s$^{-1}$)}          & \colhead{(km\,s$^{-1}$)}
}
\startdata
1 ...... & 6.7$\pm$0.5  & 11.4$\pm$0.6 & 12.2$\pm$0.7 & 33.5$\pm$2.9 & 1.30$\pm$0.01 & 1.56$\pm$0.01 \\
2 ...... & 7.8$\pm$0.8  & 11.5$\pm$0.5 & 12.3$\pm$0.6 & 21.1$\pm$2.2 & 0.95$\pm$0.01 & 1.21$\pm$0.01 \\
3 ...... & 6.5$\pm$0.8  & 11.7$\pm$0.7 & 12.5$\pm$0.9 & 18.3$\pm$2.5 & 0.87$\pm$0.01 & 0.98$\pm$0.02 \\
4 ...... & 6.9$\pm$0.8  & 10.5$\pm$0.7 & 11.1$\pm$0.9 & 17.5$\pm$2.2 & 1.06$\pm$0.01 & 1.19$\pm$0.02 \\
5 ...... & 6.6$\pm$0.6  & 11.2$\pm$0.5 & 11.9$\pm$0.6 & 15.3$\pm$1.6 & 1.01$\pm$0.01 & 1.05$\pm$0.02 \\
6 ...... & 5.6$\pm$0.8  & 11.5$\pm$0.8 & 12.4$\pm$1.0 & 16.1$\pm$2.2 & 1.21$\pm$0.02 & 1.49$\pm$0.03 \\
\enddata
\tablecomments{Columns are peak number, excitation temperature, NH$_3$ rotational temperature, kinetic temperature, NH$_3$ column density derived from the optical depth, the mean line width and the line width of the mean spectrum of the regions.}
\tablenotetext{a}{NH$_3$ rotational temperature given by \citet{ho83} as\\
\begin{displaymath}
T_{\rm rot}=-41.5 \div \ln \left\{ {-0.282 \over \tau_m(1,1)} \ln \left[ 1-{T_R^*(2,2,m) \over T_R^*(1,1,m)} \times (1-e^{-\tau_m(1,1)}) \right] \right\}.
\end{displaymath}}
\tablenotetext{b}{NH$_3$ kinetic temperature
$T_{\rm kin}={T_{\rm rot} \over 1-{T_{\rm rot} \over 42}\ln[1+1.1\exp(-16/T_{\rm rot})]}$.}
\tablenotetext{c}{NH$_3$ column density derived from its optical depth via the relation given by \citet{bac87} as\\
\begin{displaymath}
N({\rm NH_3})=2.784 \times 10^{13} \tau J(T_{\rm ex}) \Delta V\times Q/Q_1 ,
\end{displaymath}
where $Q$ is the partition function and
$Q/Q_1={1\over3}e^{23.4/T_{\rm rot}}+1+{5\over3}e^{-41.5/T_{\rm rot}}+{14\over3}e^{-101.5/T_{\rm rot}}+...$.}
\tablenotetext{d}{The mean line width is the fitting error-weighted average line width over the core regions.}
\tablenotetext{e}{The line width of the mean spectrum is the line width of mean spectrum of the spectra inside the core regions.}
\end{deluxetable}

\clearpage

\begin{deluxetable}{lcrrrrr}
\tabletypesize{\scriptsize}
\tablecolumns{7}
\tablewidth{0pc}
\tablecaption{CO and HCO$^+$ line parameters at the positions of the NH$_3$ cores.
\label{tbl:co result}}
\tablehead{
\colhead{Peak}               & \colhead{Molecule}         & \colhead{$V_{\rm LSR}$}        &
\colhead{$T^*_{\rm R}$}      & \colhead{$\Delta V$}           & \colhead{$\int{T_R^* d V}$}    &
\colhead{$\sigma$}
\\
\colhead{}                   & \colhead{}                 & \colhead{(km\,s$^{-1}$)}       &
\colhead{(K)}                & \colhead{(km\,s$^{-1}$)}   & \colhead{(K\,km\,s$^{-1}$)}    &
\colhead{(K)}
}
\startdata
1 ...... & $^{12}$CO (1$-$0) &  2.31 & 10.76 & 6.38 & 89.21$\pm$0.44 & 0.20 \\
         & $^{13}$CO (1$-$0) &  0.96 &  8.00 & 3.19 & 31.44$\pm$0.22 & 0.22 \\
         & C$^{18}$O (1$-$0) &  0.83 &  3.04 & 2.17 &  6.41$\pm$0.10 & 0.17 \\
         & HCO$^+$ (1$-$0)   &  1.44 &  1.00 & 4.76 &  4.99$\pm$0.25 & 0.21 \\
2 ...... & $^{12}$CO (1$-$0) &  2.67 & 11.02 & 6.47 & 89.02$\pm$0.72 & 0.33 \\
         & $^{13}$CO (1$-$0) &  1.03 &  7.48 & 3.08 & 30.84$\pm$0.22 & 0.22 \\
         & C$^{18}$O (1$-$0) &  1.04 &  2.75 & 1.83 &  5.21$\pm$0.13 & 0.21 \\
         & HCO$^+$ (1$-$0)   &  1.41 &  1.61 & 3.11 &  5.83$\pm$0.20 & 0.17 \\
3 ...... & $^{12}$CO (1$-$0) &  2.21 &  9.86 & 7.24 & 95.16$\pm$0.51 & 0.23 \\
         & $^{13}$CO (1$-$0) &  0.90 &  5.87 & 4.05 & 28.53$\pm$0.25 & 0.25 \\
         & C$^{18}$O (1$-$0) &  1.08 &  2.51 & 1.72 &  4.49$\pm$0.15 & 0.24 \\
         & HCO$^+$ (1$-$0)   &$-$0.52&  1.08 & 2.11 &  2.44$\pm$0.24 & 0.20 \\
4 ...... & $^{12}$CO (1$-$0) &  2.64 &  9.79 & 6.69 & 85.58$\pm$0.44 & 0.20 \\
         & $^{13}$CO (1$-$0) &  1.11 &  5.71 & 3.30 & 25.15$\pm$0.21 & 0.20 \\
         & C$^{18}$O (1$-$0) &  1.04 &  2.83 & 1.62 &  4.80$\pm$0.10 & 0.15 \\
5 ...... & $^{12}$CO (1$-$0) &  2.80 &  9.76 & 6.34 & 77.12$\pm$0.43 & 0.20 \\
         & $^{13}$CO (1$-$0) &  1.32 &  5.57 & 3.14 & 23.28$\pm$0.21 & 0.21 \\
         & C$^{18}$O (1$-$0) &  1.24 &  2.47 & 1.50 &  3.99$\pm$0.11 & 0.17 \\
6 ...... & $^{12}$CO (1$-$0) &  2.47 & 10.89 & 5.85 & 77.47$\pm$0.45 & 0.20 \\
         & $^{13}$CO (1$-$0) &  1.67 &  8.34 & 2.70 & 28.82$\pm$0.23 & 0.23 \\
         & C$^{18}$O (1$-$0) &  1.57 &  2.62 & 2.08 &  5.18$\pm$0.10 & 0.16 \\
         & HCO$^+$ (1$-$0)   &  2.08 &  1.35 & 3.33 &  4.91$\pm$0.19 & 0.16 \\
\enddata
\tablecomments{$^{12}$CO, $^{13}$CO, C$^{18}$O and HCO$^+$ line parameters are calculated within the velocity range from -3.5 to 6.8 km\,s$^{-1}$, -2.7 to 7.7 km\,s$^{-1}$, -1.0 to 3.0\,km s$^{-1}$ and 0.0 to 9.0\,km s$^{-1}$, respectively. The HCO$^+$ line was not observed towards cores P4 and P5 due to low signal-to-noise ratio.}
\end{deluxetable}

\clearpage

\begin{deluxetable}{lrcrrcrr}
\tabletypesize{\scriptsize}
\tablecolumns{8}
\tablewidth{0pc}
\tablecaption{Physical properties derived from CO.
\label{tbl:co property}}
\tablehead{
\colhead{} &
\colhead{$\rm ^{12}CO$} & \colhead{} &
\multicolumn{2}{c}{$\rm ^{13}CO$} & \colhead{} &
\multicolumn{2}{c}{$\rm C^{18}O$}
\\
\cline{4-5} \cline{7-8}
\colhead{Peak} &
\colhead{$T_{\rm ex}$} & \colhead{} &
\colhead{$\tau$} & \colhead{$N$} & \colhead{} &
\colhead{$\tau$} & \colhead{$N$}
\\
\colhead{}    &
\colhead{(K)} & \colhead{} &
\colhead{}    & \colhead{($10^{15}$cm$^{-2}$)} & \colhead{} &
\colhead{}    & \colhead{($10^{15}$cm$^{-2}$)}
}
\startdata
1 ...... & 14.17$\pm$0.20 & & 1.26$\pm$0.09 & 58.36$\pm$2.12 & & 0.35$\pm$0.03 & 8.91$\pm$0.31 \\
2 ...... & 14.43$\pm$0.33 & & 1.18$\pm$0.09 & 56.27$\pm$2.32 & & 0.31$\pm$0.03 & 6.35$\pm$0.33 \\
3 ...... & 13.26$\pm$0.22 & & 0.90$\pm$0.07 & 45.31$\pm$1.55 & & 0.34$\pm$0.03 & 6.37$\pm$0.32 \\
4 ...... & 12.90$\pm$0.19 & & 0.91$\pm$0.06 & 39.87$\pm$1.18 & & 0.39$\pm$0.03 & 6.63$\pm$0.30 \\
5 ...... & 12.25$\pm$0.20 & & 1.01$\pm$0.08 & 37.02$\pm$1.29 & & 0.33$\pm$0.03 & 4.70$\pm$0.26 \\
6 ...... & 14.30$\pm$0.21 & & 1.35$\pm$0.10 & 54.75$\pm$2.19 & & 0.30$\pm$0.02 & 7.88$\pm$0.29 \\
\enddata
\tablecomments{First column is core number. Column densities are calculated under the assumption of local thermodynamic equilibrium  (LTE), in which $^{13}$CO and C$^{18}$O have the same excitation temperature as that of optically thick $^{12}$CO.}
\end{deluxetable}

\clearpage

\begin{deluxetable}{lccccrc}
\tabletypesize{\scriptsize}
\tablecolumns{7}
\tablewidth{0pc}
\tablecaption{Velocity gradients detected in the cores\label{tbl:vgradient}}
\tablehead{
\colhead{Core}  & \colhead{Molecule}  & \colhead{$v_0\pm\sigma_{v_0}$}  &
\colhead{$\mathscr{G}\pm\sigma_\mathscr{G}$}  & \colhead{$\mathscr{G}$ at 580~pc}  & \colhead{$\Theta_\mathscr{G}\pm\sigma_{\Theta_\mathscr{G}}$} & \colhead{$\beta$}
\\
\colhead{}      & \colhead{}          & \colhead{(km\,s$^{-1}$)}   &
\colhead{(m\,s$^{-1}$\,arcsec$^{-1}$)}      & \colhead{(km\,s$^{-1}$\,pc$^{-1}$)}  & \colhead{(deg E of N)} & \colhead{}
}
\startdata
1 ...... & $^{13}$CO (1$-$0) & 1.01$\pm$0.003 & \phn4.25$\pm$0.096 & 1.51 &   $-$8.3$\pm$0.8\phn & \\
         & C$^{18}$O (1$-$0) & 0.91$\pm$0.007 & \phn3.69$\pm$0.205 & 1.31 &  $-$12.1$\pm$2.0\phn & \\
         & NH$_3$ (1,1)      & 1.15$\pm$0.014 &    10.76$\pm$0.339 & 3.83 &  $-$37.2$\pm$1.2\phn & 1.3E-2  \\
         & NH$_3$ (2,2)      & 1.28$\pm$0.024 & \phn8.76$\pm$0.927 & 3.12 &  $-$50.5$\pm$4.4\phn & \\
3 ...... & NH$_3$ (1,1)      & 1.19$\pm$0.012 & \phn3.69$\pm$0.337 & 1.31 &    165.0$\pm$4.0\phn & 2.3E-3 \\
4 ...... & NH$_3$ (1,1)      & 1.01$\pm$0.011 & \phn4.32$\pm$0.419 & 1.54 &  $-$64.5$\pm$4.0\phn & 3.6E-3 \\
5 ...... & NH$_3$ (1,1)      & 1.12$\pm$0.013 & \phn2.31$\pm$0.497 & 0.82 & $-$123.8$\pm$8.5\phn & 6.2E-4 \\
6 ...... & $^{13}$CO (1$-$0) & 1.66$\pm$0.004 & \phn2.87$\pm$0.188 & 1.02 &     16.5$\pm$2.2\phn & \\
         & C$^{18}$O (1$-$0) & 1.54$\pm$0.010 & \phn4.19$\pm$0.507 & 1.49 &     12.6$\pm$4.5\phn & \\
         & NH$_3$ (1,1)      & 1.81$\pm$0.018 & \phn9.45$\pm$0.772 & 3.36 &     90.2$\pm$3.9\phn & 1.5E-2 \\
         & NH$_3$ (2,2)      & 1.88$\pm$0.048 &    14.92$\pm$2.772 & 5.31 &     45.6$\pm$7.6\phn & \\
\enddata
\tablecomments{Columns are core number, molecules fitted, systemic velocity, magnitude of the velocity gradient, $\mathscr{G}$ at cloud distance, direction of increasing velocity (measured east of north), and parameter $\beta$ given in \S\ref{sec:rot}. Errors quoted are $1~\sigma$ uncertainties.}
\end{deluxetable}

\clearpage

\begin{deluxetable}{lccccc}
\tabletypesize{\scriptsize}
\tablecolumns{6}
\tablewidth{0pc}
\tablecaption{NH$_3$ abundance in the cores\label{tbl:abundance}}
\tablehead{
\colhead{Peak}       &
\colhead{$N(\rm H_2)_{C^{18}O}$}   & \colhead{$\chi_{\rm {NH}_3}$} &
\colhead{$N(\rm H_2)_{1.12mm}$}                     & \colhead{$\chi'_{\rm {NH}_3}$}&
\colhead{$\chi~{\rm in~core}$}
\\
\colhead{}                        &
\colhead{($10^{22}\rm cm^{-2}$)}  & \colhead{($10^{-8}$)} &
\colhead{($10^{22}\rm cm^{-2}$)}  & \colhead{($10^{-8}$)} &
\colhead{($10^{-8}$)}
}
\startdata
1 ...... & 6.23 & 5.4 & 5.36 & 6.3  & 2.41 \\
2 ...... & 4.45 & 4.7 & 5.54 & 3.8  & 2.74 \\
3 ...... & 4.46 & 4.1 & 2.85 & 6.4  & 2.46 \\
4 ...... & 4.64 & 3.8 & 2.16 & 8.1  & 2.24 \\
5 ...... & 3.29 & 4.7 & 0.96 & 15.9 & 1.88 \\
6 ...... & 5.52 & 2.9 & 1.25 & 12.9 & 2.04 \\
\enddata
\tablecomments{The columns show the core number, H$_2$ column density and NH$_3$ abundance $\chi$ based on C$^{18}$O observations (column 2 and 3) and 1.1~mm dust flux (column 4 and 5), and the average NH$_3$ abundance over the cores.}
\end{deluxetable}

\clearpage

\begin{deluxetable}{lccccccc}
\tabletypesize{\scriptsize}
\tablecolumns{8}
\tablewidth{0pc}
\tablecaption{Masses of the NH$_3$ cores\label{tbl:core}}
\tablehead{
\colhead{Region}        & \colhead{R} &
\colhead{$n({\rm H_2})$}                 &
\colhead{$M_{\rm Jeans}$}                & \colhead{$M_{\rm Virial}$} &
\colhead{$M_{\rm Molecular}$} & \colhead{$\alpha$} &
\colhead{$v_{\rm escape}$}
\\
\colhead{}                               & \colhead{(arcsec)} &
\colhead{($10^{4}\rm cm^{-3}$)}          &
\colhead{($M_\sun$)}                     & \colhead{($M_\sun$)} &
\colhead{($M_\sun$)}                     & \colhead{} &
\colhead{(km\,s$^{-1}$)}
}
\startdata
1 ...... &  69.0 & 13.8 & 198 & 206 & 208 & 0.99 & 3.03 \\
2 ...... &  74.9 & 13.2 & 205 & 121 & 255 & 0.41 & 3.22 \\
3 ...... &  69.0 & ~8.7 & 259 & ~92 & 130 & 0.71 & 2.40 \\
4 ...... &  60.1 & 10.4 & 198 & 119 & 104 & 1.15 & 2.30 \\
5 ...... &  46.7 & 16.6 & 174 & ~85 & ~77 & 1.10 & 2.24 \\
6 ...... &  49.7 & 11.0 & 228 & 129 & ~62 & 2.08 & 1.95 \\
\enddata
\tablecomments{Columns are core number, size of the cores, H$_2$ densities derived from C$^{18}$O, Jeans mass, virial masses from NH$_3$ and gas mass, virial parameter $\alpha$, and escape velocity.}
\end{deluxetable}

\end{document}